\documentclass[aps,prd,preprintnumbers,showpacs,showkeys,nofootinbib,
superscriptaddress,fleqn,floatfix,tightenlines,10pt]{revtex4-1}
\usepackage{amsmath,amsfonts,amssymb,amscd,amsxtra,amsthm}
\usepackage{graphicx}  
\usepackage{epstopdf}
\usepackage{dcolumn}  
\usepackage{bm}        
\usepackage{slashed}
\usepackage{cancel} 
\usepackage{float}
\usepackage{mathtools}
\usepackage{amsbsy}
\usepackage{amstext}

\usepackage[normalem]{ulem}
\usepackage[dvipsnames]{xcolor}

\usepackage[utf8]{inputenc} 
\usepackage{booktabs}
\usepackage[normalem]{ulem} 
\usepackage[dvipsnames]{xcolor} 
\usepackage{tabularx}
\usepackage{enumitem}  
\usepackage{array} 
\usepackage{multirow}
\renewcommand\sout{\bgroup \color{red} \ULdepth=-.5ex \ULset}

\makeatletter

\begin{document}  
\preprint{INHA-NTG-03/2021}
\title{\Large Production of hidden-charm strange pentaquarks $P_{cs}$
  from the $K^- \,p \to J/\psi \,\Lambda$ reaction} 

\author{Samson Clymton}
\email[E-mail: ]{sclymton@inha.edu}
\affiliation{Department of Physics, Inha University,
Incheon 22212, Republic of Korea }

\author{Hee-Jin Kim}
\email[E-mail: ]{heejin.kim@inha.edu}
\affiliation{Department of Physics, Inha University,
Incheon 22212, Republic of Korea }

\author{Hyun-Chul Kim}
\email[E-mail: ]{hchkim@inha.ac.kr}
\affiliation{Department of Physics, Inha University,
Incheon 22212, Republic of Korea }
\affiliation{School of Physics, Korea Institute for Advanced Study 
  (KIAS), Seoul 02455, Republic of Korea}
\begin{abstract}
We investigate the production of the hidden-charm pentaquark
$P_{cs}^0(4459)$ with strangeness in the $K^- p \to J/\psi \Lambda$
reaction, employing two different theoretical frameworks, i.e., the
effective Lagrangian method and the Regge approach. Having determined
all relevant coupling constants, we are able to compute the total
and differential cross sections for the $K^- p \to J/\psi \Lambda$
reaction. We examine the contributions of $P_{cs}$ with different sets
of spin-parity quantum number assigned. The present results may give a
guide for possible future experiments. 
\end{abstract}
\maketitle
\section{Introduction}
Very recently, the LHCb Collaboration has announced the finding of a
new hidden-charm pentaquark state with strangeness in the analysis of
$\Xi_b^-\to J/\psi\Lambda K^-$ 
decays~\cite{Aaij:2020gdg}. This hidden-charm pentaquark baryon with 
strangeness  is christened as $P_{cs}^0(4459)$. The mass and width of
$P_{cs}$ is determined to be respectively $4458.8\pm
2.9_{-1.1}^{+4.7}$ MeV and $17.3_{-5.7}^{+8.0}$ MeV. While the quark 
content of $P_{cs}^0(4459)$ can be given as $udsc\bar{c}$, its
spin-parity quantum number is not known yet because of lack of the
data. This finding broadens our understanding of how the quarks form  
multi-quark hadrons in addition to the heavy pentaquark baryons 
$P_c$~\cite{Aaij:2015tga, Aaij:2016phn, Aaij:2019vzc} and
many charmonium-like tetraquark mesons~\cite{Choi:2003ue,
  Aubert:2003fg} (see recent experimental and theoretical
reviews~\cite{Chen:2016qju, Esposito:2016noz, Dong:2017gaw,
  Olsen:2017bmm, Guo:2017jvc}). The structure of $P_c$ and $P_{cs}$
has been theoretically studied in various works~
\cite{Maiani:2015vwa,
  Li:2015gta, Ghosh:2017fwg, Cheng:2015cca,
 Anisovich:2015zqa, Wang:2015wsa, Chen:2015sxa, Feijoo:2015kts,
 Lu:2016roh, Chen:2016ryt, Xiao:2019gjd, Wang:2019nvm, Chen:2020uif,
 Peng:2020hql, Chen:2021tip,
 Wu:2010jy, Wu:2010vk, Yuan:2012wz, Santopinto:2016pkp, Takeuchi:2016ejt,
 Yamaguchi:2016ote, Yamaguchi:2017zmn, Yamaguchi:2019seo, He:2019ify}.
The internal structure of the hidden-charm 
pentaquark states is still under debate. Since the mass of
$P_{cs}^0(4459)$ is about 19 MeV below the $\bar{D}^* \Xi_c^0$
threshold, it is arguably considered to be a hadronic molecular
state~\cite{Chen:2015sxa,Chen:2016ryt, Xiao:2019gjd, Wang:2019nvm,
  Chen:2020uif, Peng:2020hql, Chen:2021tip}. On the other 
hand, the hidden-charm pentaquark states are interpreted as compact
pentaquarks consisting of two diquarks and an antiquark bound
states~\cite{Maiani:2015vwa, Li:2015gta, Wang:2015wsa, Ghosh:2017fwg, 
Wang:2020eep}, hadrocharmonium states~\cite{Eides:2019tgv,Anwar:2018bpu,
Ferretti:2020ewe}, coupled-channel unitary approach with the local 
hidden gauge formalism~\cite{Wu:2010jy, Wu:2010vk}, five-quark states
~\cite{Yuan:2012wz, Santopinto:2016pkp, Takeuchi:2016ejt}, 
meson-baryon molecules with coupled channels~\cite{Yamaguchi:2016ote},
meson-baryon molecules coupled to the five-quark states
~\cite{Yamaguchi:2017zmn, Yamaguchi:2019seo}, and as hadronic molecule 
states in a quasi-potential Bethe-Salpeter equation approach
~\cite{He:2019ify}.
Theoretically, the
spin-parity quantum number of the $P_{cs}^0(4459)$ is proposed to be
$1/2^-(3/2^-)$. Reference~\cite{Peng:2020hql} argues that $J^P=3/2^-$
is preferable over $J^P=1/2^-$ based on the hadronic molecular picture 
of $P_{cs}^0(4459)$, though it should be determined by experiments.    
   
In principle, the hidden-charm pentaquark states can be produced by
meson beams such as the pion and kaon. Since several experimental
programs to measure charmed hadrons have been planned at the Japan
Proton Accelerator Research Complex (J-PARC)~\cite{Noumi, Shirotori,
  Kim:2014qha, Kim:2015ita, Kim:2016imp}, it is also of great
importance to investigate the production mechanism of 
the hidden-charm pentaquark states. In Ref.~\cite{Lu:2015fva},
the production of the $P_c^0(4380)$ and $P_c^0(4450)$ was studied 
in the $\pi^- p\to J/\psi n$ reaction, based
on the effective Lagrangian approach. This approach provides a simple but
clear understanding of how the $P_c$'s can be created at the level of
the Born approximation. The transition amplitude includes the $P_{c}$'s 
as the resonance baryons in the $s$ channel explicitly together
with $\pi$ and $\rho$ exchanges in the $t$ channel and the $P_c$'s
exchange  in the $u$ channel. They found that the contributions of the 
$P_c^0(4380)$ and $P_c^0(4450)$ bring about the clear peak structures
in order of $1\mu\mathrm{b}$ at the energies corresponding to the
masses of $P_c$'s. On the other hand, Ref.~\cite{Kim:2016cxr} examined
the $\pi^- p\to J/\psi n$ reaction, using the Regge approach. The
$t$ channel for the hidden charm reaction is distinguished from that
for the open charm reaction, since the hidden charm processes are
suppressed by the Okubo-Zweig-Iizuka (OZI) rule. This indicates that
it is difficult to determine the coupling constant for $P_c$ by using
some model calculations. Thus, one needs to make a reasonable
assumption for the branching ratios of $P_c$. A similar situation is
expected also for the $K^- p\to J/\psi \Lambda$ reaction.

In the present work, we investigate the production of $P_{cs}^0(4459)$
in the $K^-p\to J/\psi \Lambda$ reaction, based on two different
theoretical models, i.e. the effective Lagrangian method and Regge
approach. In particular, since the energy of the initial kaon should
be enough to create the charmonium $J/\psi$ and $\Lambda$, it is
worthwhile to consider also the Regge approach. In
Refs.~\cite{Kim:2014qha, Kim:2015ita}, both the effective Lagrangian
method and Regge approach were used for the study of the open-charm
process $\pi^- p\to D^{*-}\Lambda_c^+$. It turns out that the Regge
approach describes the experimental data very well over the whole
energy region. However, while the Regge approach describes the general
behavior of the cross sections at very high energies, it has certain
difficulties to describe experimental data quantitatively. One
effective way of improving this Regge approach is that one can replace
the Feynman propagators in the transition amplitudes derived based on
the effective Lagrangian by the Reggeized propagator. This method is
often called the hybridized Regge approach. 
Actually, the Regge approach was used for the description of the
$\pi^- p\to J/\psi n $ reaction~\cite{Kodaira:1979sf} in which the
total cross section for the reaction was estimated to be around $1$ pb
at the momentum $p=50\,\mathrm{GeV}/c$. Moreover,  
the hybridized Regge approach was developed and successfully applied 
to photoproduction of mesons~\cite{Guidal:1997hy}.  
In the present work, we take the same strategy such that we will
employ both the effective Lagrangian and Regge approaches and compare
the results each other, since these two approaches are complementary
each other. Since the
spin-parity quantum number  
of $P_{cs}^0(4459)$ is experimentally unknown, we will consider six
different cases, i.e. $J^{P}=1/2^{\pm}$, $J^{P}=3/2^{\pm}$, and
$J^{P}=5/2^{\pm}$, emphasizing the cases of $J^P=1/2^-$ and $3/2^-$.
Then, we scrutinize the differences among the contributions of
$P_{cs}$ to the $K^- p \to J/\psi \Lambda$ with the different 
spin-parity quantum number assigned. 
The present work will provide helpful guidance on possible
future experiments at the J-PARC and on determining the spin-parity
quantum number of $P_{cs}$. 

We sketch the present work as follows: In Section II, we explain the
general formalism for the effective Lagrangian and Regge
approaches. Since the coupling constants at the vertices including
$P_{cs}$ are not known, we first estimate them by imposing reasonable
assumptions on the branching ratios of the $P_{cs}$ decays. 
In Section III, we present the results for the total and differential
cross sections, emphasizing the differences arising from different
spin-parity quantum numbers. In the final Section, we summarize the
present work and will draw conclusions.

\section{General formalism}
We first start with the effective Lagrangian approach and then will
continue to formulate the transition amplitude for the $K^-p\to J/\psi
\Lambda$ reaction in the Regge approach.
\subsection{Effective Lagrangian method}
\begin{figure}[htp]
\includegraphics[scale=1.2]{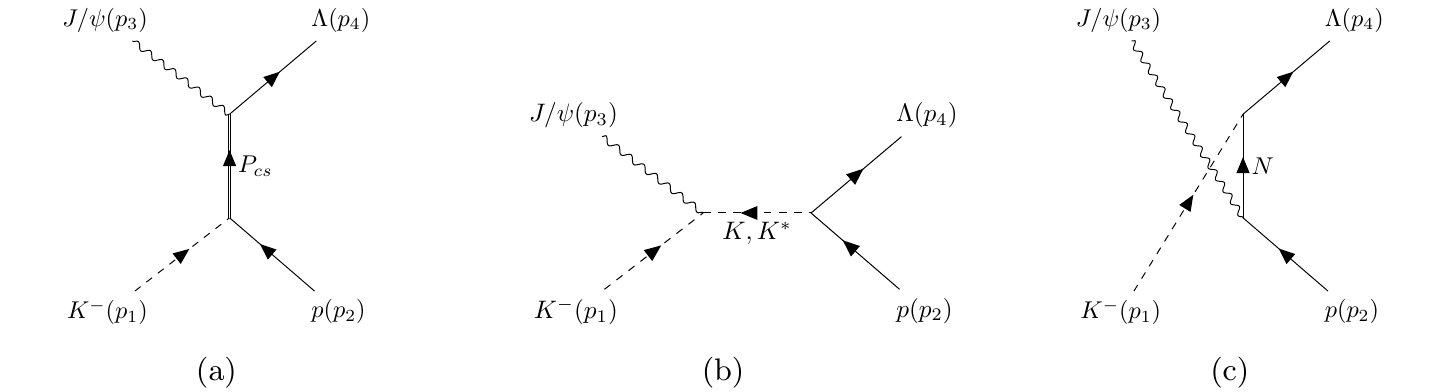}
\caption{The tree-level Feynman diagrams for the $K^- + p\to J/\psi +
  \Lambda$ reaction. In the left panel the $s$-channel is drawn,
  whereas in the center and right panels, the $t$-channel and
  $u$-channel diagrams are depicted. $p_i$ stand for the four-momenta
  of hadrons involved in the reaction.} 
\label{fig:feydiag}
\end{figure}
In the effective Lagrangian approach for the $K^-p\to J/\psi
\Lambda$ reaction, we can consider three different Feynman diagrams
that are drawn in Fig.~\ref{fig:feydiag}. In the $s$ channel, we can only
include $P_{cs}^0(4459)$ with the experimental data on its mass and
decay width taken into account~\cite{Aaij:2020gdg}. Though we can
include other hyperons with strangeness $S=-1$, we will neglect them,
because we do not have any information on the coupling constant for
the vertices such as $Y\Lambda J/\psi$ and furthermore their
contributions will be negligible, since they are far from
on-mass-shell. The $t$ channel contains $K$ and $K^*$ exchange. In the
$u$ channel, we can introduce the nucleon. Note that it is not
possible to include $P_{cs}$ in the $u$ channel, which is very
different from the case of the $\pi^- p \to J/\psi n$ reaction.  
Since the spin-parity quantum number of $P_{cs}^0(4459)$ is unknown, 
we assume six different cases: $J^{P}=1/2^\pm,\,3/2^\pm,\,5/2^\pm$.
Taking into account these different quantum numbers, we can express the
effective Lagrangians for $P_{cs}$ as follows~\cite{Lu:2015fva,
Kim:2016cxr,Mart:2015jof,Kim:2011rm,Wang:2015jsa}
\begin{align}
\mathcal{L}_{P\Lambda J/\psi}^{1/2\pm}=&\, -g_{P\Lambda J/\psi}
\bar{P}\Gamma^{\mp}_\mu\Lambda\psi^{\mu} 
+ \frac{f_{P\Lambda J/\psi}}{2 m_\Lambda}
 \bar{P}\sigma_{\mu\nu}\Gamma^{\pm}\Lambda\psi^{\mu\nu}
 + \mathrm{h.c.},\cr 
\mathcal{L}_{P\Lambda J/\psi}^{3/2\pm}=&\,-\frac{g_{P\Lambda
 J/\psi}}{2 m_\Lambda}  \bar{P}_{\mu}\Gamma^{\pm}_\nu\Lambda\psi^{\mu\nu}
-\frac{f_{P\Lambda J/\psi}}{4 m_\Lambda^2}
 \bar{P}_{\mu}\Gamma^{\mp}\partial_{\nu} 
\Lambda\psi^{\mu\nu} -\frac{h_{P\Lambda J/\psi}}{4 m_\Lambda^2}
 \bar{P}_{\mu}\Gamma^{\mp}\Lambda\partial_{\nu}\psi^{\mu\nu}+
 \mathrm{h.c.},\cr  
\mathcal{L}_{P\Lambda J/\psi}^{5/2\pm}=&\, 
-\frac{g_{P\Lambda J/\psi}}{2 m_\Lambda^2}
 \bar{P}_{\mu\alpha}\Gamma^{\mp}_\nu\Lambda\partial^\alpha 
\psi^{\mu\nu}-\frac{f_{P\Lambda J/\psi}}{4 m_\Lambda^3} \bar{P}_{\mu\alpha} 
\Gamma^{\pm}\partial_{\nu}\Lambda\partial^\alpha\psi^{\mu\nu} 
-\frac{h_{P\Lambda J/\psi}}{4 m_\Lambda^3} \bar{P}_{\mu\alpha}\Gamma^{\pm} 
\Lambda\partial^\alpha\partial_{\nu}\psi^{\mu\nu} +\mathrm{h.c.},
\label{Eq:pljcoup}
\end{align}
where $P$, $\Lambda$, $\psi^\mu$ denote the fields corresponding
respectively to $P_{cs}^0(4459)$, $\Lambda^0$, and
$J/\psi$. $\psi_{\mu\nu}$ is defined as $\partial_\mu \psi_\nu
-\partial_\nu \psi_\mu$. $m_\Lambda$ stands for the mass of the
$\Lambda$ hyperon. $\Gamma_\mu$ and $\Gamma$ are given respectively by   
\begin{align}
\Gamma^{\pm}_\mu = 
\begin{pmatrix} 
\gamma_\mu\gamma_5 \\ \gamma_\mu 
\end{pmatrix} 
\mbox{ and }
\Gamma^{\pm} = \begin{pmatrix} 1 \\ i\gamma_5
\end{pmatrix},
\end{align}
with different parities considered. Since we consider the production
of $P_{cs}$ in the vicinity of the $J/\psi \Lambda$ threshold, we will
take into account the first terms in each effective Lagrangians.
We will consider only the terms with $g_{P\Lambda J/\psi}$ in the
effective Lagrangian, assuming that those with $f_{P\Lambda J/\psi}$
and $h_{P\Lambda J/\psi}$ are rather small near the threshold. 

The effective Lagrangians for the $NP_{cs}K$ vertex are written as 
\begin{align}
\mathcal{L}_{P N K}^{1/2\pm}=&\,- g_{PNK} \bar{P}
\Gamma^{\mp}N K +  \mathrm{h.c.},\cr 
\mathcal{L}_{P N K}^{3/2\pm}=&\,- \frac{g_{PNK}}{M_{P_{cs}}\, m_N} 
\varepsilon^{\mu\nu\alpha\beta}\partial_\mu\bar{P}_\nu 
\Gamma^{\pm}_\alpha N \partial_\beta K + \mathrm{h.c.},\cr
\mathcal{L}_{P N K}^{5/2\pm}=&\,\, - \frac{g_{PNK}}{M_{P_{cs}} \,m_N^2} 
\varepsilon^{\mu\nu\alpha\beta}\partial_\mu\bar{P}_{\nu\rho} 
\Gamma^{\mp}_\alpha N \partial^\rho\partial_\beta K 
+ \mathrm{h.c.},
\end{align}
where $M_{P_{cs}}$ and $m_N$ represent the masses of $P_{cs}$ and the
nucleon respectively. 

Since there is no information on the coupling constants for the 
$P_{cs}J/\psi \Lambda$ and $P_{cs}KN$ vertices experimentally, 
it is very difficult to determine them. As will be discussed
soon, one possible way is to resort to some guessworks based
on theoretical works~\cite{Kim:2016cxr,Paryev:2018fyv,Wang:2019dsi}
and recent experimental data on $\pi N$ and $\bar{K}N$
scattering~\cite{Jenkins:1977xb,Chiang:1986gn,Zyla:2020zbs}. 
Note that we have used the $\pi N$ and $\bar{K}N$ scattering 
data to extrapolate the $P_{cs}J/\psi\Lambda$ and $P_{cs}KN$ coupling
constants. This is an assumption justified by the fact that the energy
of the $P_{cs}$ production is rather high such that the effects of the
explicit SU(3) symmetry breaking are also suppressed, considering the 
fact that the ratio between the strange current quark
mass $m_s$ and the kinetic energy of the $\Lambda$ baryon,
is rather small. The coupling constants for $P_{cs}$ are
extracted by using the partial-wave decay width given by    
\begin{align}
\Gamma (P_{cs}\to M B) =& \frac{|\mathbf{k}|}{8\pi
  M_{P_{cs}}^2}\,\frac{1}{2J+1} \sum_{\lambda_1 = -J}^{J}
  \sum_{\lambda_2,\lambda_3} |A(P_{cs}\to M B)|^2 ,
\label{eq:decaywidth}
\end{align}
where $M$ and $B$ denote the produced meson and baryon in the final
state, respectively. $|\mathbf{k}|$ is the momentum of the meson in
the final state and $J$ represents the total angular momentum of the final
state. The $\lambda_i$ are the spin projections of the particles
involved. The decay amplitudes $A(P_{cs}\to MB)$ for $P_{cs}\to
J/\psi\Lambda$ are obtained from the effective Lagrangian with
spin-parity quantum numbers for $P_{cs}$ given 
\begin{align}
A^{1/2\pm}_{P \Lambda J/\psi} =&\, -g_{P\Lambda J/\psi}\,
\bar{u}_{P}\,\Gamma^{\mp}_\mu\, \epsilon^{\mu}\, u_\Lambda, \cr
A^{3/2\pm}_{P \Lambda J/\psi} =&\, i\frac{g_{P\Lambda J/\psi}}{2m_\Lambda}\,
\bar{u}_{P\mu}\,\Gamma^{\pm}_\nu\, (q_\psi^\mu\epsilon^\nu -
  q_\psi^\nu\epsilon^\mu)\,u_\Lambda ,  \cr
A^{5/2\pm}_{P \Lambda J/\psi} =&\, \frac{g_{P\Lambda J/\psi}}{2m_\Lambda^2}\,
\bar{u}_{P\mu\alpha}\,\Gamma^{\mp}_\nu\, 
(q_\psi^\mu\epsilon^\nu
 -q_\psi^\nu\epsilon^\mu)\,q_\psi^\alpha\,u_\Lambda, 
\label{eq:5-7}
\end{align}
whereas those for $P_{cs}\to KN$ are expressed as
\begin{align}
A^{1/2\pm}_{P N K} =& \,- g_{PNK}\, \bar{u}_{P} \,\Gamma^{\mp}\,
                      u_{N},\cr 
A^{3/2\pm}_{P N K} =& \,- \frac{g_{PNK}}{M_{P_{cs}}\, m_N} 
\varepsilon_{\mu\nu\alpha\beta} \,\bar{u}_P^\nu 
\,q_P^\mu\,\Gamma_{\pm}^\alpha\, q_{K}^\beta\, u_{N},\cr
A^{5/2\pm}_{P N K} =& \, i\frac{g_{PNK}}{M_{P_{cs}}\, m_N^2} 
\varepsilon_{\mu\nu\alpha\beta}\,\bar{u}_P^{\nu\rho}\,q_P^\mu\, 
\Gamma_{\mp}^\alpha\, q_{K}^\beta \,q_{K \rho}\, u_{N}.
\end{align}
$\epsilon_\mu$ in Eq.~\eqref{eq:5-7} stands for the polarization vector of
$J/\psi$. $q_i^\mu$ ($i=P,K,\psi$) denote respectively the momenta of
$P_{cs}$, $K$ and $J/\psi$ in the center of mass(CM) frame. Note that
$P_{cs}$ is at rest before it decays. The Rarita-Schwinger spinor for
$P_{cs}$ with higher spins ($s\geq 3/2$) is given by 
the following recursive equation~\cite{Rarita:1941mf} 
\begin{align}
u^{n+1/2}_{\mu_1\cdots\mu_{n-1}\mu}(p,s) \equiv \sum_{r,m}
  \left(n+1/2,s|1,r;n-1/2,m\right)
  u^{n-1/2}_{\mu_1\cdots\mu_{n-1}}(p,m) \varepsilon_{\mu}^{r}(p), 
\end{align}
where $s, m$ and $r$ designate the projections of spin-$(n+1/2)$,
spin-$(n-1/2)$, and the polarization of a massive spin-$1$ particle 
respectively.  

To determine the coupling constants for the $P_{cs}J/\psi \Lambda$ and
$P_{cs} KN$ vertices, one should know the experimental data on their
branching ratios. Unfortunately, however, they are not known at all.
Even in the case of the $P_c$ its branching ratios are 
unknown experimentally. This means that we have to make reasonable
assumptions of the branching ratios of $P_{cs}\to J/\psi\Lambda$ and
$P_{cs}\to KN$.  A previous investigation on photoproduction of the
hidden-charm pentaquark $P_c(4450)$~\cite{Paryev:2018fyv} proposed
that if the branching ratio of $P_c(4450)\to J/\psi p$ is $1\%$ or
less, then one can explain the threshold enhancement of the $J/\psi$ 
production due to $P_{c}$ and the modification of the $J/\psi$ mass in
nuclear medium. However, this is still a very crude estimate for the 
branching ratio of $P_c \to J/\psi p$. Note that even the $\pi^- p\to
J/\psi n$ reaction was not much studied experimentally and only the
upper limit of its total cross section is
known~\cite{Jenkins:1977xb, Chiang:1986gn}. Nevertheless, in 
Refs.~\cite{Kim:2016cxr, Wang:2019dsi}, the upper limit of the total 
cross section for the $\pi^- p\to J/\psi n$ reaction was cautiously
investigated with $P_c$ resonances taken into account, in which the
constraint on the branching ratio of  $P_c$ was discussed, especially
in the region near the threshold.  The branching ratio of $P_c \to
J/\psi n$ decay was estimated to be about a few percents whereas $P_c
\to \pi^- p$ decay was given to be of order  $10^{-4}$, since it is
the OZI-suppressed process. These estimates are in agreement with  
recent findings from the GlueX Collaboration~\cite{Ali:2019lzf}.

When it comes to that for $P_{cs}$ decays, the situation is even
worse than the $P_c$ case. Since there is no experimental information
on the decay of $P_{cs}$ at all, it is very difficult to determine the
coupling constant for the $P_{cs}J/\psi\Lambda$ and $P_{cs} K N$
vertex. Nevertheless, it is 
worthwhile to estimate the branching ratio of the $P_{cs}\to J/\psi
\Lambda$. Since the threshold energy of the $P_{cs}$ production is
rather high, the effects of the explicit SU(3) symmetry breaking are
also suppressed, considering the fact that the ratio between the
strange current quark mass $m_{\mathrm{s}}$ and the kinetic energy of
$\Lambda$, $T_K(\Lambda)$, is rather small
($m_{\mathrm{s}}/T_K(\Lambda)\ll 1$). Actually, this assumption is a
reasonable one, since the magnitude of the total cross section of $K^-
p $ scattering is similar to $\pi^- p $
scattering~\cite{Zyla:2020zbs}. Based on this assumption, we are 
able to estimate the upper limit of the total cross section for the
$K^- p\to J/\psi \Lambda$ reaction near threshold to be around 1
nb. This implies that the branching ratios of the $P_{cs} \to 
J/\psi\Lambda$ and $K^- p$ decays are about  $1\%$ and $0.01\%$
respectively. If the branching ratio of $P_{cs} \to J/\psi\Lambda$
were larger than $10~\%$, then one would have found the evidence for
the existence of $P_{cs}$ already from the old data for $K^- p $
scattering, which we will discuss later. Moreover, note 
that this $1~\%$ branching ratio of the $P_{cs} \to J/\psi\Lambda$
decay is in line with recent investigations on the structure of
$P_{cs}$ with the molecular picture taken into
account~\cite{Chen:2016qju, Xiao:2021rgp}. 

Using this estimate of the branching ratio, we can obtain the coupling 
constant for the $P_{cs}KN$ vertex. The results for the coupling constants 
for $P_{cs}$ are listed in Table~\ref{tab:1}. Note that we take the positive values
for the coupling constants. 
\setlength{\tabcolsep}{5pt}
\renewcommand{\arraystretch}{1.5}
\begin{table}[htp]
\caption{Numerical results for the coupling constants 
$g_{P_{cs}J/\psi \Lambda}$ and $g_{P_{cs} KN}$. The branching ratios
of $P_{cs}\to J/\psi\Lambda$ and $P_{cs} \to pK$ decays are assumed to be
$1\,\%$ and $0.01\,\%$, respectively. Note that we choose the
positive values for the coupling constants.} 
\label{tab:1}
\begin{tabular}{l c c c c c c}
\hline 
    \hline 
$g_{P_{cs}MB}(J^P)$ & $1/2^+$ & $1/2^-$ & $3/2^+$ & $3/2^-$ & $5/2^+$
  & $5/2^-$ \\ 
\hline
$P_{cs}\,J/\psi\,\Lambda$ & $1.26\times 10^{-1}$ & 
$4.41\times 10^{-2}$ & $1.48\times 10^{-1}$ & $5.46\times 10^{-2}$ 
& $1.33\times 10^{-1}$ & $3.83\times 10^{-1}$ \\
$P_{cs}\,K\,p$            & $5.82\times 10^{-3}$ 
& $3.77\times 10^{-3}$ & $2.06\times 10^{-3}$ 
& $3.18\times 10^{-3}$ & $1.84\times 10^{-3}$ 
& $1.19\times 10^{-3}$ \\
\hline 
    \hline 
\end{tabular}%
\end{table}

Once the values of the coupling constants are given, it is
straightforward to express the transition amplitudes in the 
$s$ channel
\begin{align}
\mathcal{M}_{1/2^\pm} &= i g_{P\Lambda J/\psi} g_{PNK}\, 
\bar{u}(p_4,\lambda_4)
\,\Gamma^\mu_\mp \,\epsilon_\mu^*(p_3,\lambda_3)\, 
\frac{\slashed{q} + M_{P_{cs}}}{s - M_{P_{cs}}^2} \,
\Gamma_\mp  \,u(p_2,\lambda_2),
\label{eq:8} \\
\mathcal{M}_{3/2^\pm} &= -\frac{g_{P\Lambda J/\psi}
                        g_{PNK}}{2M_{P_{cs}} 
m_N m_\Lambda} \bar{u}(p_4,\lambda_4)\, \Gamma_\nu^\pm \,(p_3^\mu 
\epsilon^{*\nu}(p_3,\lambda_3) - \epsilon^{*\mu}(p_3,\lambda_3)
                        p_3^\nu) 
\frac{\Delta_{\mu\sigma}}{s - M_{P_{cs}}^2}\, \varepsilon^{\rho \sigma
                        \alpha \beta} 
q_\rho \,\Gamma_\alpha^\pm \, p_{1\beta}\, u(p_2,\lambda_2), 
\label{eq:9}\\ 
\mathcal{M}_{5/2^\pm} &= -\frac{g_{P\Lambda J/\psi}
                        g_{PNK}}{2M_{P_{cs}} 
m_N m_\Lambda^2} \bar{u}(p_4,\lambda_4)\, \Gamma_\nu^\mp p_3^{\lambda}
                        \,(p_3^\mu 
\epsilon^{*\nu}(p_3,\lambda_3) - \epsilon^{*\mu}(p_3,\lambda_3)
                        p_3^\nu) 
\frac{\Delta_{\mu \lambda \sigma \delta}}{s - M_{P_{cs}}^2}
                        \,\varepsilon^{\rho \sigma 
\alpha \beta} q_\rho \,\Gamma_\alpha^\mp\, p_{1\beta}\,  p_1^\delta \,
                        u(p_2,\lambda_2), 
\label{eq:10}
\end{align}
where $\epsilon^*_\mu$ denotes the polarization vector for $J/\psi$
and $q$ stands for the momentum of $P_{cs}$ given by 
$q=p_1+p_2=p_3+p_4$. Taking into account the decay width of $P_{cs}$,
we change the $P_{cs}$ mass $M_{P_{cs}}$ in the propagator to be
$(M_{P_{cs}} -i\Gamma_{P_{cs}}/2)$. The spin projection operators for
$P_{cs}$ with spin 3/2 and 5/2 are defined respectively
as~\cite{Kim:2012pz} 
\begin{align} \label{}
\Delta_{\mu \sigma} &= (\slashed{q} + M_{P_{cs}}) 
\left[-g_{\mu \sigma} + \frac{1}{3} \gamma_\mu \gamma_\sigma 
+ \frac{1}{3M_{P_{cs}}}(\gamma_\mu q_\sigma - \gamma_\sigma q_\mu)
+ \frac{2}{3M_{P_{cs}}^2} q_\mu q_\sigma \right], \cr
\Delta_{\mu \lambda \sigma \delta} &= (\slashed{q} + M_{P_{cs}}) 
\left[ \frac{1}{2} (\bar{g}_{\mu \sigma} \bar{g}_{\lambda \delta} +
\bar{g}_{\mu \delta} \bar{g}_{\lambda \sigma})
- \frac{1}{5} \bar{g}_{\mu \lambda} \bar{g}_{\sigma \delta}
- \frac{1}{10} (\bar{\gamma}_\mu \bar{\gamma}_\sigma \bar{g}_{\lambda \delta}
+ \bar{\gamma}_\mu \bar{\gamma}_\delta \bar{g}_{\lambda \sigma}
+ \bar{\gamma}_\lambda \bar{\gamma}_\sigma \bar{g}_{\mu \delta}
+ \bar{\gamma}_\lambda \bar{\gamma}_\delta \bar{g}_{\mu \sigma}) \right],
\end{align}
where
\begin{align} \label{}
\bar{g}_{\mu \nu} = g_{\mu \nu} - \frac{q_\mu q_\nu}{M_{P_{cs}}^2}, \;\;\;
\bar{\gamma}_\mu = \gamma_\mu - \frac{q_\mu}{M_{P_{cs}}^2} \slashed{q}.
\end{align}

In the $t$-channel, we consider the exchange of the $K$ and $K^*$
mesons. The effective Lagrangians for the $J/\psi KK$ and $J/\psi
KK^*$ vertices are given as  
\begin{align}
\mathcal{L}_{J/\psi K K} =&\,\, -i g_{J/\psi KK}\, \psi^\mu
                            \left(K^+\partial_\mu K^- -
                            K^-\partial_\mu K^+\right),\cr
\mathcal{L}_{J/\psi K K^*} =&\,\, -\frac{g_{J/\psi KK^*}}{m_{\psi}}
                              \varepsilon^{\mu\nu\alpha\beta}\partial_\mu\psi_\nu 
                              K \partial_\alpha K^*_\beta,
\label{eq:elback}
\end{align}
where $m_{\psi}$ denotes the mass of $J/\psi$. The coupling constant will
be determined by using a similar method as in the $s$-channel case. The
decay amplitudes for the corresponding decays in Eq.~\eqref{eq:elback}
are obtained to be 
\begin{align}
A_{J/\psi K K} =& \,- g_{J/\psi KK}\,(q_K-q'_K)_\mu \epsilon^\mu ,\cr
A_{J/\psi K K^*} =&\, -\frac{g_{J/\psi KK^*}}{m_{\psi}}
                    \varepsilon^{\mu\nu\alpha\beta}q_{\psi\mu}
                    \,q_{K^*\alpha}\,\epsilon_\nu\,
                    \epsilon^*_{K^*\beta} , 
\end{align}
where $q'_K$ stands for the momentum of the kaon that goes to the
opposite direction with $q_K$. The polarization vector of $K^*$ is
expressed by $\epsilon^\mu_{K^*}$. Since the decay widths of
$J/\psi$ to the $K$ and $K^*$ mesons are experimentally known
as~\cite{Zyla:2020zbs}  
\begin{align}
\Gamma_{J/\psi\to KK} =&\, 2.66\times 10^{-2}\,\mathrm{keV},\;\;\;
                         \Gamma_{J/\psi\to KK^*} = 5.57\times
                         10^{-1}\,\mathrm{keV}, 
\end{align}
we can directly obtain the coupling constants $g_{J/\psi KK}$ and
$g_{J/\psi KK^*}$, respectively, as follows
\begin{align}
g_{J/\psi KK} =&\, 7.12\times 10^{-4}, \;\;\; g_{J/\psi KK^*} =
                 8.82\times 10^{-3}. 
\label{eq:16}
\end{align}
Those for the $\Lambda N K$ and $\Lambda N K^*$ vertices are rather
well known. The semi-phenomenological nucleon-hyperon interaction such
as the Nijmegen extended-soft-core model
(ESC08a)~\cite{Rijken:2010zzb} provides us with their values. Then,
the effective Lagrangians for the $\Lambda N K$ and $\Lambda N K^*$
vertices are expressed as 
\begin{align}
\mathcal{L}_{\Lambda N K} =&\,\, -\frac{f_{\Lambda N
                             K}}{m_\pi}\bar{\Lambda} \gamma_\mu
                             \gamma_5 N \partial^\mu K +
                             \mathrm{h.c.},\\ 
\mathcal{L}_{\Lambda N K^*} =&\,\, -g_{\Lambda N K^*}
                               \bar{\Lambda}\gamma^{\mu} N K^*_\mu -
                               \frac{f_{\Lambda N K^*}}{4
                               m_N}\bar{\Lambda}\sigma^{\mu\nu} N
                               \left(\partial_\mu K^*_\nu
                               - \partial_\nu K^*_\mu\right) +
                               \mathrm{h.c.}, 
\end{align}
with the coupling constants given by
\begin{align}
f_{\Lambda N K} = -0.2643,\;\;\; g_{\Lambda N K^*} = -1.1983,\;\;\;
 f_{\Lambda N K^*} = -4.2386\,. 
\end{align}
Thus, the resulting transition amplitudes for $K$ and $K^*$ exchanges
are respectively given as 
\begin{align}
\mathcal{M}_{K} =& \, \frac{g_{J/\psi KK}f_{\Lambda N
                   K}}{m_\pi}\,\bar{u}(p_4,\lambda_4)\gamma_5 \frac{(2
                   p_1 -
                   p_3)\cdot\epsilon^{*}(p_3,\lambda_3)}{t-m_K^2}
                   \slashed{q}_t u(p_2,\lambda_2),\\ 
\mathcal{M}_{K^*} =& \, i  \frac{g_{J/\psi KK^*}g_{\Lambda N
                     K^*}}{m_\psi}\,\bar{u}(p_4,\lambda_4)
                     \frac{\varepsilon_{\mu\nu\alpha\beta}
                     p_3^\mu\epsilon^{*\nu}(p_3,\lambda_3)
                     q_t^\alpha}{t-m_{K^*}^2}
                     \left(-g^{\beta\sigma}+\frac{q_t^\beta    
                     q_t^\sigma}{m_{K^*}^2}\right) \left( 
                     \gamma_\sigma +
                     i\frac{\kappa_{K^*}}{2m_N}
                     \sigma_{\gamma\sigma}q_t^\gamma\right) 
                     u(p_2,\lambda_2),
\end{align}
where $q_t = p_3-p_1$  and $\kappa_{K^*}=f_{\Lambda N
  K^*}/g_{\Lambda N K^*}$. 

As for the $u$-channel contribution, we consider only the $N$
exchange. The effective Lagrangian for the $NNJ/\psi$ vertex is
similar to the $P_{cs}$ with spin-$1/2^+$ as in Eq.~\eqref{Eq:pljcoup}
\begin{align}
\mathcal{L}_{J/\psi N N}=&\, -g_{J/\psi NN}\bar{N}\gamma_\mu
                           \psi^{\mu}N - \frac{f_{J/\psi NN}}{2M_N}
                           \bar{N}  \sigma_{\mu\nu}\psi^{\mu\nu} N
 + \mathrm{h.c.}.
\end{align}
Since the $J/\psi$ vector meson has a nature similar to the $\phi$
vector meson, we ignore the second term with the tensor coupling
constant, since its value is related to the charmed magnetic moment of
the nucleon, which can be neglected. It is also difficult to determine
the vector coupling constant $g_{J/\psi NN}$. We take its value from
Ref.~\cite{Barnes:2006ck}:
$g_{J/\psi NN}=g_{J/\psi N\bar{N}}=1.62\times10^{-3}$. This small
value indicates already that the $u$-channel contribution will be very
tiny.  The corresponding $u$-channel amplitude is obtained as 
\begin{align}
\mathcal{M}_N = -\frac{g_{J/\psi NN}f_{\Lambda N
                 K}}{m_\pi}\,\bar{u}(p_4,\lambda_4)\gamma_5
                 \slashed{p}_1\,\frac{\slashed{q}_u + m_N}{u - m_N^2}
                 \, \slashed{\epsilon}^*(p_3,\lambda_3)
                 u(p_2,\lambda_2),  
\end{align}
where $q_u = p_4-p_1$.

Since hadrons have finite sizes and structures, it is essential to
consider a form factor at each vertex. 
Actually, there is no firm theoretical ground as to
how one can determine the values of the cutoff masses. In
practice, the values of the cutoff masses are usually
fitted to the experimental data. Unfortunately, we do not
have experimental data enough to determine them
in the present case. Nevertheless, there is one
theoretical guideline. As discussed in Ref.~\cite{Kim:2018nqf},
heavier baryons are considered to be more compact than lighter ones, 
which was found by examining the electromagnetic form
factors of singly heavy baryons. By "\textit{more compact}"
we mean that the intrinsic size of the heavier baryons (or
hadrons) should be smaller than the light ones, which leads
to larger values of the cutoff masses in general. Being
guided by this, we have chosen the cutoff masses $\Lambda$
in such a way that $\Lambda - m \simeq 600-700$ MeV.
In the present work, we will take the 
form factors, which are most used in reaction calculations. So, 
we introduce the form factors in the $s$-, $t$- and
$u$-channels, respectively, as follows: 
\begin{align}
F_s (q^2) &= \frac{\Lambda^4}{\Lambda^4+(s-m^2)^2},\cr
F_t (q_t^2) &= \frac{\Lambda^2-m^2}{\Lambda^2-t},\cr
F_u (q_u^2) &= \frac{\Lambda^2-m^2}{\Lambda^2-u} , 
\end{align}
with the values of the cutoff masses taken to be 
\begin{align}
\Lambda_{P_{cs}} = 5.0\,\mathrm{GeV}, \;\;\;
\Lambda_K = 1.0\,\mathrm{GeV}, \;\;\;
\Lambda_{K^*} = 1.4\,\mathrm{GeV}, \;\;\; 
\Lambda_N = 1.5\,\mathrm{GeV}. 
\end{align}
Note that these values of the cutoff masses have been used in
various different reactions. 

\vspace{0.5cm}

\subsection{Regge Approach}
The effective Lagrangian method is known to describe well the hadronic 
productions at low-energy regions, in particular, in the vicinity of
the threshold energy. However, since this method is based on the Born
approximation, i.e., a tree-level calculation, it is not suitable to
explain the exclusive or diffractive hadronic processes at higher
energies. On the other hand, the Regge approach explains the general
high-energy behaviors of the hadronic reactions but only
qualitatively. To overcome this disadvantage, a hybridized Regge
approach was phenomenologically proposed~\cite{Guidal:1997hy} in an
attempt to improve the Regge approach quantitatively. This approach is
characterized by replacing the Feynman propagator derived from the
effective Lagrangian method by the Regge one
  \begin{align} \label{}
  \frac{1}{t-m_X^2} \longrightarrow \mathcal{P}_{\mathrm{Regge}}^\pm =
  -\Gamma\left(-\alpha_X
  (t)\right) \xi_X^\pm \alpha_X' \left(\frac{s}{s_0}\right)^{\alpha_X(t)}.
  \end{align}
This method was successfully applied to hadronic reactions
throughout broad energy regions including even the resonance
regions, $\sqrt{s}\sim 3~\mathrm{GeV}$~\cite{Kim:2014qha,Kim:2015ita}.  
 
\subsubsection{$K$ and $K^*$ Reggeon exchange}
\begin{figure}[htp]
\includegraphics[scale=1]{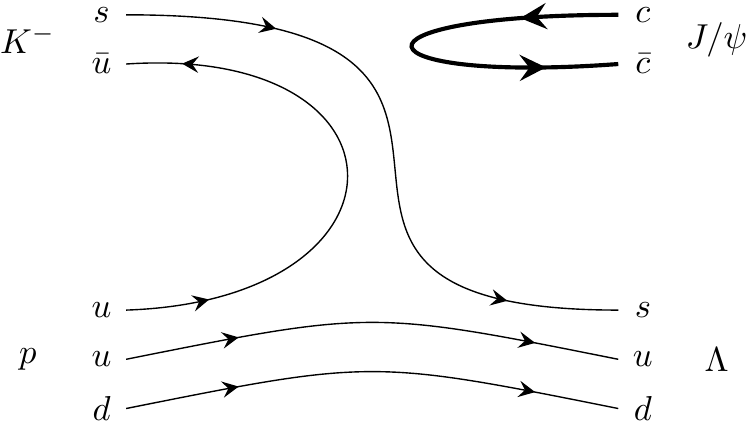}
\caption{The $t$-channel schematic diagram for the hidden-charm $K p\to
J/\psi \Lambda$ reaction.}
\label{fig:2}
\end{figure}
In Fig.~\ref{fig:2}, we depict schematically the $t$-channel diagram
in terms of the quark lines~\cite{Kim:2016cxr}. As shown in
Fig.~\ref{fig:2}, hadronic $J/\psi$ productions by the photon, $\pi$ 
or $K$ beams are all OZI suppressed, being similar to the $\phi$-meson
production. So, we consider the light-Reggeon exchanges in the
$t$-channel, i.e. the $K$ and $K^*$ Reggeons. We employ here a
hybridized Regge method, in which the Feynman propagators in the
transition amplitudes obtained in the previous subsection are replaced
by the Regge
propagator~\cite{Don:2002,Guidal:1997hy,Kim:2016imp,Kim:2017nxg}.   
Thus, we can express the transition amplitudes with the $K$- and
$K^*$-Reggeon exchanges, respectively, as  
\begin{align}
\mathcal{M}_{K}^R(s,t) &= -\mathcal{M}_{K}(s,t) 
\left\{\begin{array}{c}
1 \\
e^{-i\pi\alpha_K(t)}
\end{array}\right\}
\Gamma(-\alpha_K
(t))\alpha_K'(m_K^2)\left(\frac{s}{s_0}\right)^{\alpha_K(t)}
  \left(t-m_K^2\right), \\ 
\mathcal{M}_{K^*}^R(s,t) &= -\mathcal{M}_{K^*}(s,t) 
\left\{\begin{array}{c}
1 \\
e^{-i\pi\alpha_{K^*}(t)}
\end{array}\right\}\Gamma(1-\alpha_{K^*}
(t))\alpha_{K^*}'(m_{K^*}^2)\left(\frac{s}{s_0}\right)^{\alpha_{K^*}(t)-1}
  \left(t-m_{K^*}^2\right), 
\label{Eq:regge}
\end{align}
where $\alpha_K$ and $\alpha_{K^*}$ denote the Regge trajectories for
the $K$ and $K^*$ mesons, respectively. $\alpha'(t)$ represents the
derivative of $\alpha$ with respect to $t$: $\alpha'(t)=\partial
\alpha/ \partial t$. The scale parameter $s_0$ is a free
parameter. Though this can be fitted to the data, if they exist, its
value is widely taken to be $s_0 = 1~\mathrm{GeV}^2$, which
corresponds to a typical hadronic scale. This can be also estimated
theoretically. If the $t$-channel diagram as shown in
Fig.~\ref{fig:2} were a planar diagram, the energy-scale parameter
$s_0$ could have been calculated by using the planar diagram
decomposition~\cite{Titov:2008yf,Kim:2017hhm}. However, the
$t$-channel diagram for the $K^- p\to J/\psi \Lambda$ reaction is not
a planar one. So, there is no clear way to determine the value
of $s_0$. In the present work, we will utilize the result of Model I
as a guideline to determine $s_0$. Since the Regge amplitude have 
to be consistent with that of Model I at the Regge pole position, we
extract the value of $s_0$ by comparing the results for the $d\sigma/
dt$ from Model I with those for Model II near the pole. The reasonable
values of $s_0$ turn out to be $s_0= 5~ \mathrm{GeV}^2$ for $K^{*}$-
and $s_0 = 2~\mathrm{GeV}^2$ for 
  $K$-Reggeon exchange.

\begin{figure}[htp]
\includegraphics[scale=1]{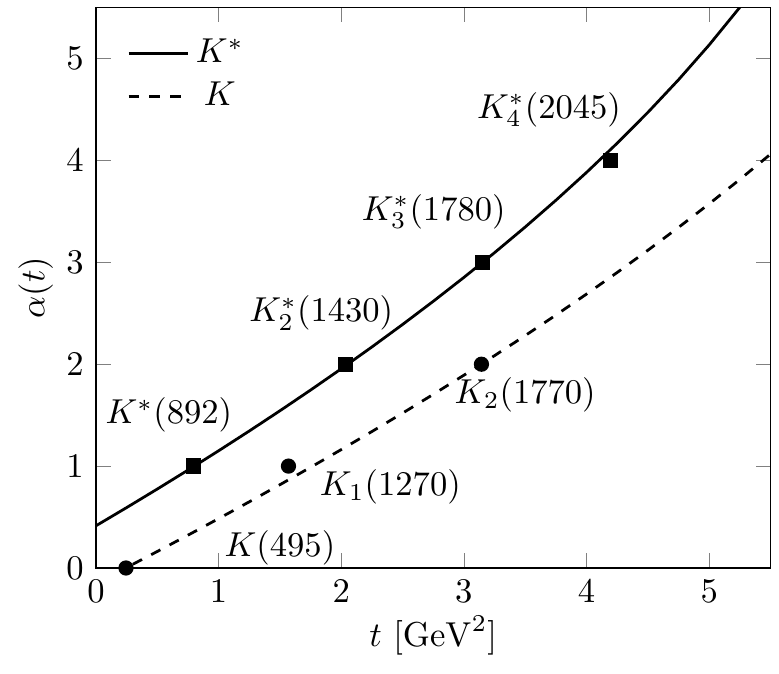}
\includegraphics[scale=1]{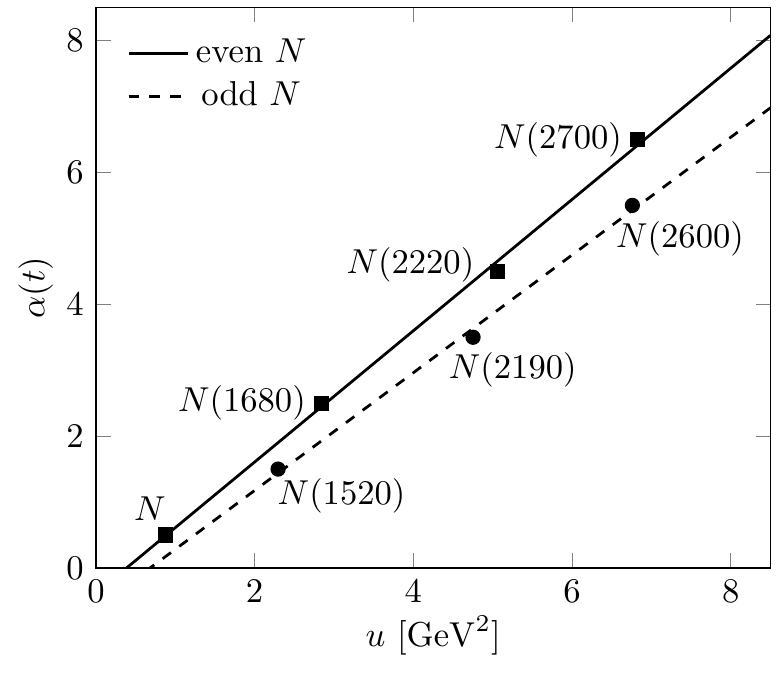}
\caption{Regge trajectories for $K$, $K^*$ and nucleon}
\label{fig:3}
\end{figure}
\begin{figure}[htp]
\includegraphics[scale=0.4]{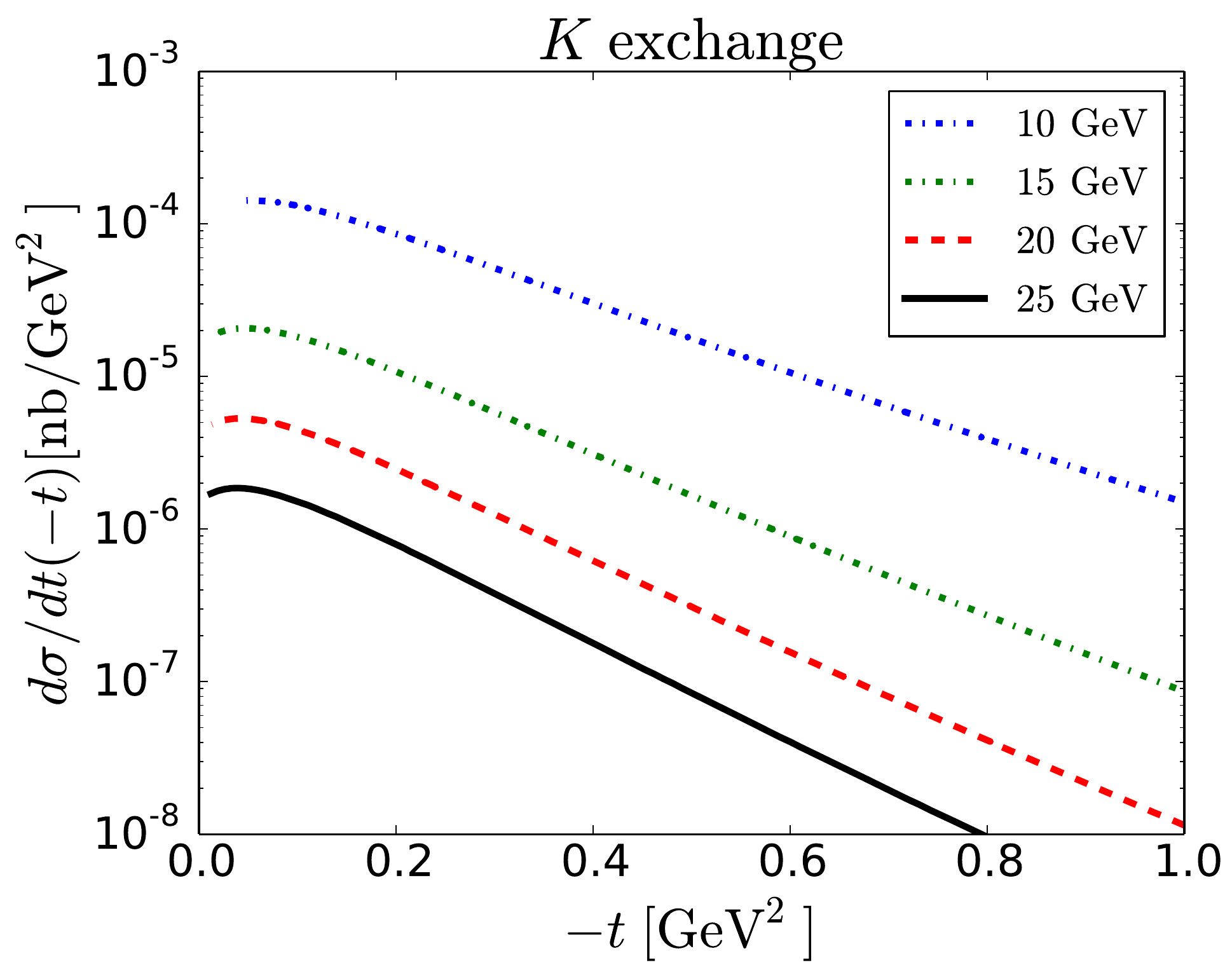}
\includegraphics[scale=0.4]{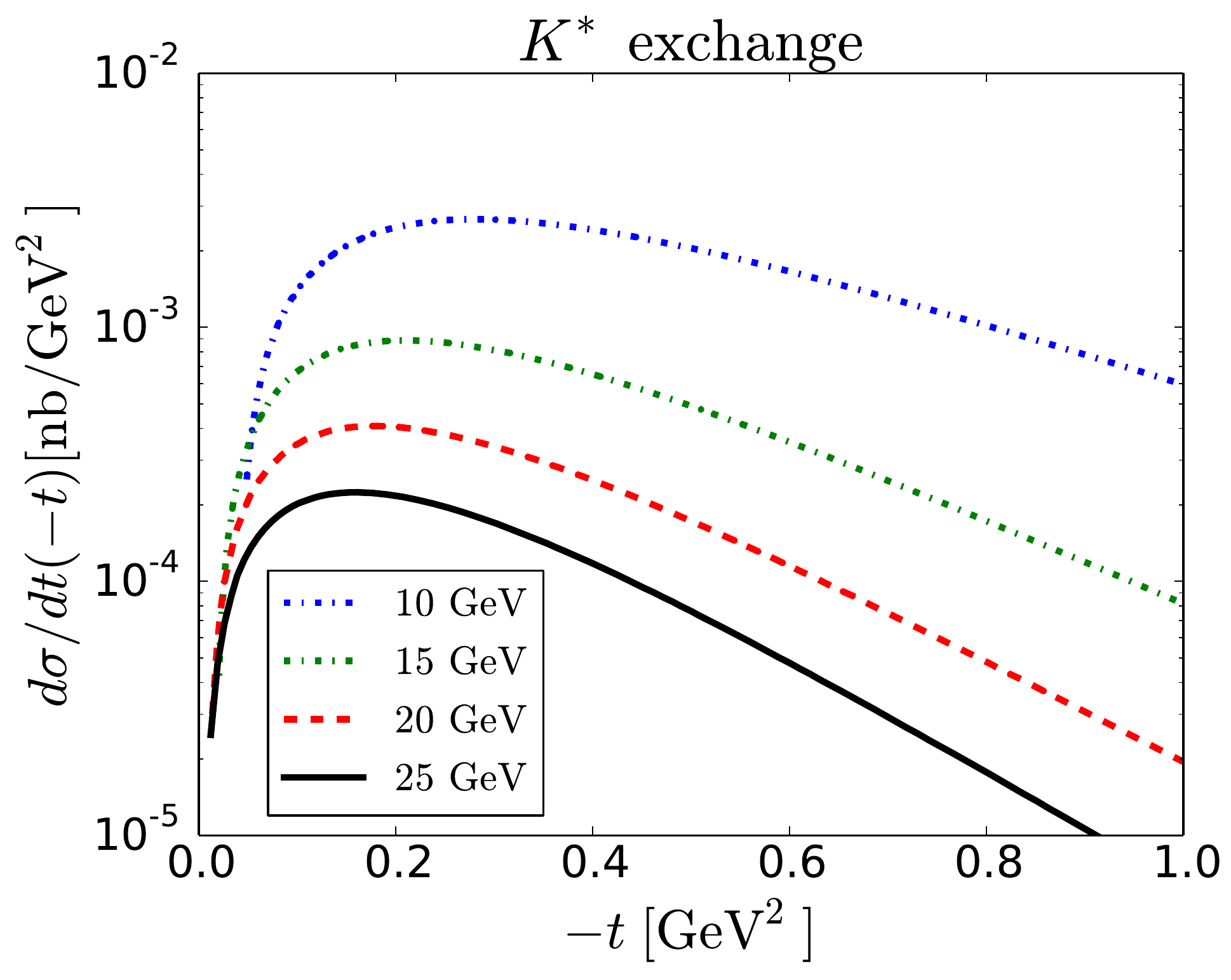}
\caption{$d\sigma/dt$ as a function of $-t$ for the $K$ and $K^*$
contributions from $W=10$ GeV to $25$ GeV.}
\label{fig:4}
\end{figure}

Though the linear Regge trajectories are given, we will adopt the
nonlinear Regge trajectories~\cite{Brisudova:1999ut}, since it
describes the trajectories more realistically as shown in  
Fig.~\ref{fig:3}. Thus, $\alpha_K$ and $\alpha_{K^*}$ are parametrized
as 
\begin{align}
\alpha_{K(K^*)}(t) = \alpha_{K(K^*)}(0) + \gamma \left(\sqrt{T_{K(K^*)}} - 
\sqrt{T_{K(K^*)}-t}\right),
\end{align}
where $\gamma$ governs the slope of the trajectories and $T_{K(K^*)}$
denote their terminal points. The parameters for the $K$ and $K^*$
trajectories are fixed to be 
\begin{align} \label{Eq:parameters}
&\gamma = 3.65~\mathrm{GeV}^{-1}, \;\;\;
\alpha_K(0) = -0.151,\;\;\; \alpha_{K^*}(0) = 0.414, \cr
&\sqrt{T_{K}} = 2.96~\mathrm{GeV}, \;\;\; \sqrt{T_{K^*}} =
  2.58~\mathrm{GeV}. 
\end{align}
Note that in the limit $t \to 0$, this square-root trajectory reduces
to the linear function
\begin{align} \label{eq:linear_regge}
\alpha(t) \approx \alpha(0) + \frac{\gamma}{2\sqrt{T}}t 
= \alpha(0) + \alpha'(0) t.
\end{align}
Before we carry out the numerical calculation, it is of great interest
to examine the asymptotic behavior of the differential cross section
$d\sigma/dt$. It is known that in the large $s$ limit the asymptotic
behavior of $d\sigma/dt$ is given as   
\begin{align} \label{eq:asymp_s}
\frac{d\sigma}{dt}(s \to \infty, t \to 0) \propto s^{2\alpha(0)-2}.
\end{align}
We found that the transition amplitudes are proportional to $t$
and $s$ as follows
\begin{align} \label{eq:aymp_feynamp}
\lim_{s\to\infty} \sum_{\lambda_i, \lambda_f}
  \left|\mathcal{M}_{K^*}\left(t-m_{K^*}^2\right)\right|^2 \propto 
s^2t
\end{align}
and the differential cross section
\begin{align} 
\frac{d\sigma}{dt} = \frac{1}{64\pi s} \frac{1}{|p_\mathrm{cm}|^2} 
\sum_{\lambda_i,\lambda_f} \left|\mathcal{M}_{K^*}^R\right|^2 
\propto \sum_{\lambda_i,\lambda_f} \left|\mathcal{M}_{K^*}
\left(t-m_{K^*}^2\right)\right|^2 
s^{2\alpha(t) - 4} \underset{s\to\infty}{\propto} s^{2\alpha(t)-2},
\label{eq:dcs}
\end{align}
which reproduces correctly the asymptotic behavior given in
Eq.~\eqref{eq:asymp_s}. Here, $p_\mathrm{cm}$ stands for the initial
momentum in the CM frame, which is proportional to $\sqrt{s}$ in the
large $s$ limit. The numerical results for $d\sigma/dt$ with $K$ and
$K^*$ considered only are depicted in Fig.~\ref{fig:4}. As one can
see already in Eq.~\eqref{eq:aymp_feynamp}, the contribution of
$K^*$ exchange to $d\sigma/dt$ decreases rapidly at very
forward scattering $t \to 0$ in the same context of $\gamma N \to 
K\Lambda$~\cite{Guidal:1997hy} and $\pi N \to
K^*\Lambda$~\cite{Kim:2015ita} reactions.   
As $t$ increases, $d\sigma/dt$ falls off linearly for $K$
exchange, whereas that for $K^*$ exchange grows very fast in
the forward direction, and then decreases almost linearly.

As shown in the left panel of Fig.~\ref{fig:3}, the even and odd
signatured $K$ ($K^*$) poles are lying on the same trajectory, which
means that the $K$ $(K^*)$ Regge trajectory is \textit{degenerate}. 
When the total transition amplitudes are derived, the even and odd Regge
propagators can be added or
subtracted~\cite{Guidal:1997hy,Corthals:2005ce}.    
Thus, the Regge propagator for $K$ ($K^*$) thus contains either $1$
(constant phase) or $e^{-i\pi\alpha (t)}$ (rotating phase). However,
we find that the results for the total and differential cross
sections are not much changed by the signature factor, so we choose the
constant signature factor. On the other hand, note that the asymmetry
will be quite sensitive to this factor, which will not be computed in
the present work.  

\subsubsection{$N$ Reggeon exchange}
We will follow the same method for the nucleon Reggeon in the $u$-channel.
Replacing the Feynman propagator by the Regge propagator, we obtain
the transition amplitudes for the $u$-channel as follows 
\begin{align}
\mathcal{M}_{R}(s,u) = -\mathcal{M}_N(s,u) \xi^+_N
\Gamma(0.5-\alpha_N(u))\alpha_N'\left(\frac{s}{s_0}\right)^{\alpha_N(u)-0.5}
  \left(u-m_N^2\right).
\label{Eq:regge-u}
\end{align}
We take the linear trajectory as in Ref.~\cite{Storrow:1983ct}. Based on the
nucleon trajectory drawn in the right panel of Fig.~\ref{fig:3}, we find the
Regge trajectory for the even signatured nucleon~\cite{Zyla:2020zbs} as 
\begin{align}
\alpha_N(u) = \alpha_N(0) + \alpha_N' u\,; \;\;\; 
\alpha_N(0) = -0.384, \;\; \alpha_N' = 0.996\,.
\end{align}
Since one can distinguish the even and odd $N$ trajectory for the
nucleon, so the signature factor for the nucleon Regge trajectory can
be taken to be 
\begin{align} \label{eq:nuc_sig}
\xi^+_N = \frac{1 + e^{-i\pi \alpha_N(u)}}{2}.
\end{align}
The energy-scale parameter $s_0$ cannot be obtained by using the
similar way as in the $t$-channel because of the following reason. It
is related to the asymptotic behavior of the $u$-channel Regge
propagator. At very high energy and in the very forward direction, which
correspond to $s\to\infty$ and $t\to 0$, respectively, we get
$u\approx -s$ that leads to $\alpha(u)\approx -\alpha' s$. Moreover, using
the asymptotic behavior of the $\Gamma$ function when $z\to \infty$,
we find an approximated relation
\begin{align}
\Gamma(z)\approx \sqrt{2\pi(z-1)}\left(\frac{z-1}{e}\right)^{z-1}.
\end{align}
Thus, Eq.~\eqref{Eq:regge-u} is reduced to 
\begin{align}
\mathcal{M}_{R}(s\to \infty,u\approx -s) \approx \mathcal{M}_{F}(s,u)
  C s^\beta \left(\alpha_N' s_0/e \right)^{\alpha_N' s}. 
\label{eq:37}
\end{align}
The last factor in Eq.~\eqref{eq:37} gives a hint on $s_0$. If $\alpha'
s_0 > e$, then the above given amplitude will diverge as $s$
grows. Since $\alpha_N'$ is less than $1$, we are able to fix the
energy-scale parameter to be $s_0 = 2 \,\mathrm{GeV}^2$ such that the
amplitude is kept to be convergent. 
\vspace{0.5cm}

\section{Results and discussion}
\begin{figure}[htp]
\includegraphics[scale=0.7]{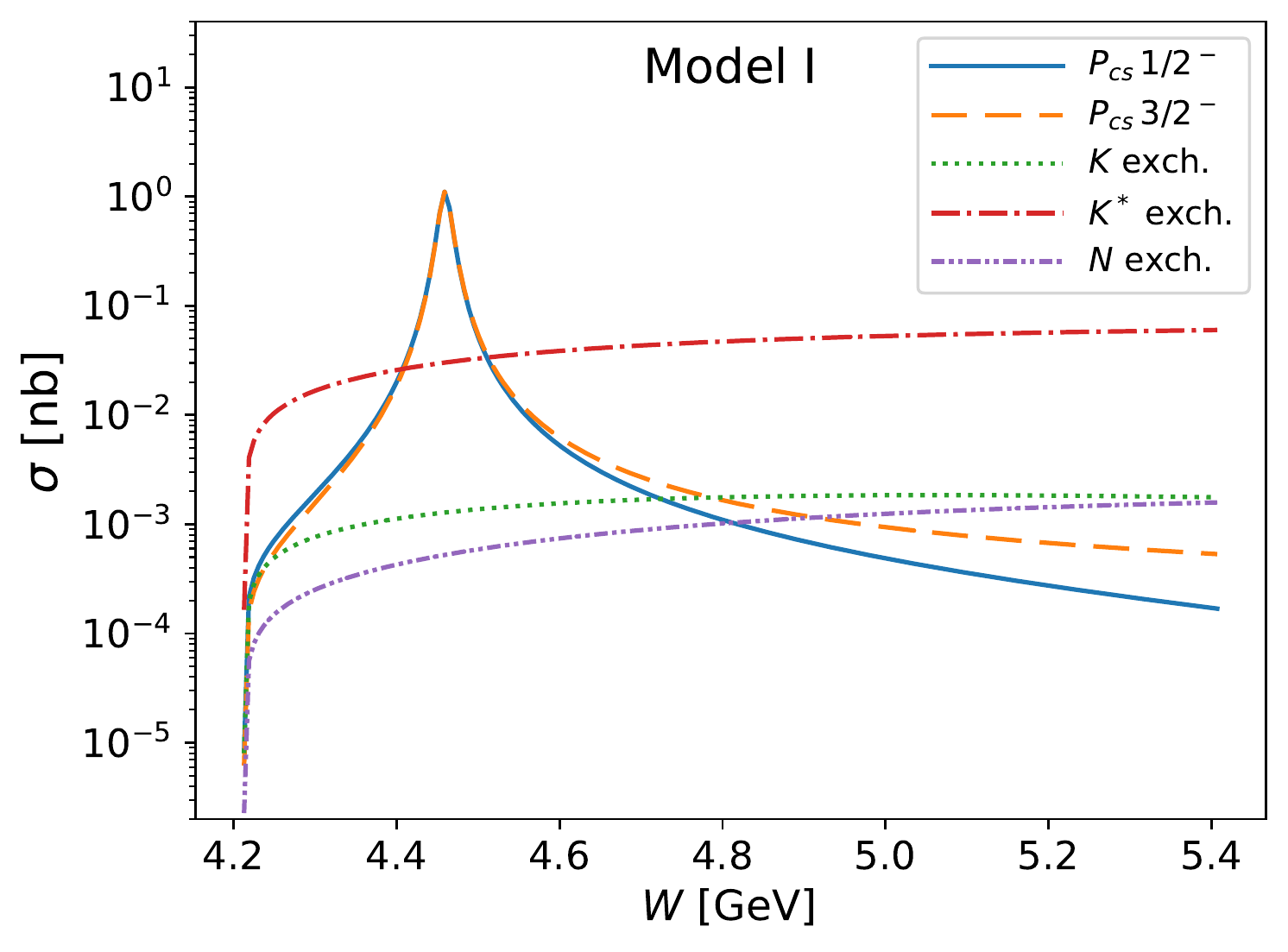}
\caption{Numerical results for the total cross section as a function
  of the total CM energy ($W$) from Model I. We consider two different 
  cases of spin-parity quantum number for $P_{cs}$, i.e. $J^P=1/2^-$
  and $J^P=3/2^-$. The $s$-channel contribution is drawn in the solid 
  and dashed curves in the case of $J^P=1/2^-$ and
  $J^P=3/2^-$, respectively. The dot-dashed curve depicts the
  contribution from $K^*$ exchange in the $t$ channel, whereas the
  dotted one illustrates that from $K$ exchange. The two-dot-dashed one
  draws the contribution from $N$ exchange in the $u$ channel.}   
\label{fig:5}
\end{figure}
We first examine each contribution to the total cross section. In
Fig.~\ref{fig:5}, we show the results for each contribution to the
total cross section for the $K^- p\to J/\psi \Lambda$ reaction. We
consider here the hidden-charm pentaquark $P_{cs}$ with $J^P=1/2^-$
and $J^P = 3/2^-$ in the $s$ channel. The resonance peak reaches the
magnitude of nb order, i.e. $\sigma \sim 1$ nb at $W\approx 4.46$
GeV. The contribution from $K^*$ exchange in the $t$ channel is the
most dominant one apart from the resonance region. Those from $K$ and
$N$ exchanges are negligibly small, since they are approximately 100
times smaller than the contribution from $K^*$ exchange. The reason
can be found from the difference in the values of the coupling
constants, given in Eq.~\eqref{eq:16}. The coupling constant for the
$J/\psi KK^*$ vertex is at least ten times larger than that for the
$J/\psi KK$ vertex. Thus, the contribution from $K^*$ exchange to the
total cross section is much larger than those from both $K$ and $N$
exchanges. 

\begin{figure}[htp]
\includegraphics[scale=0.7]{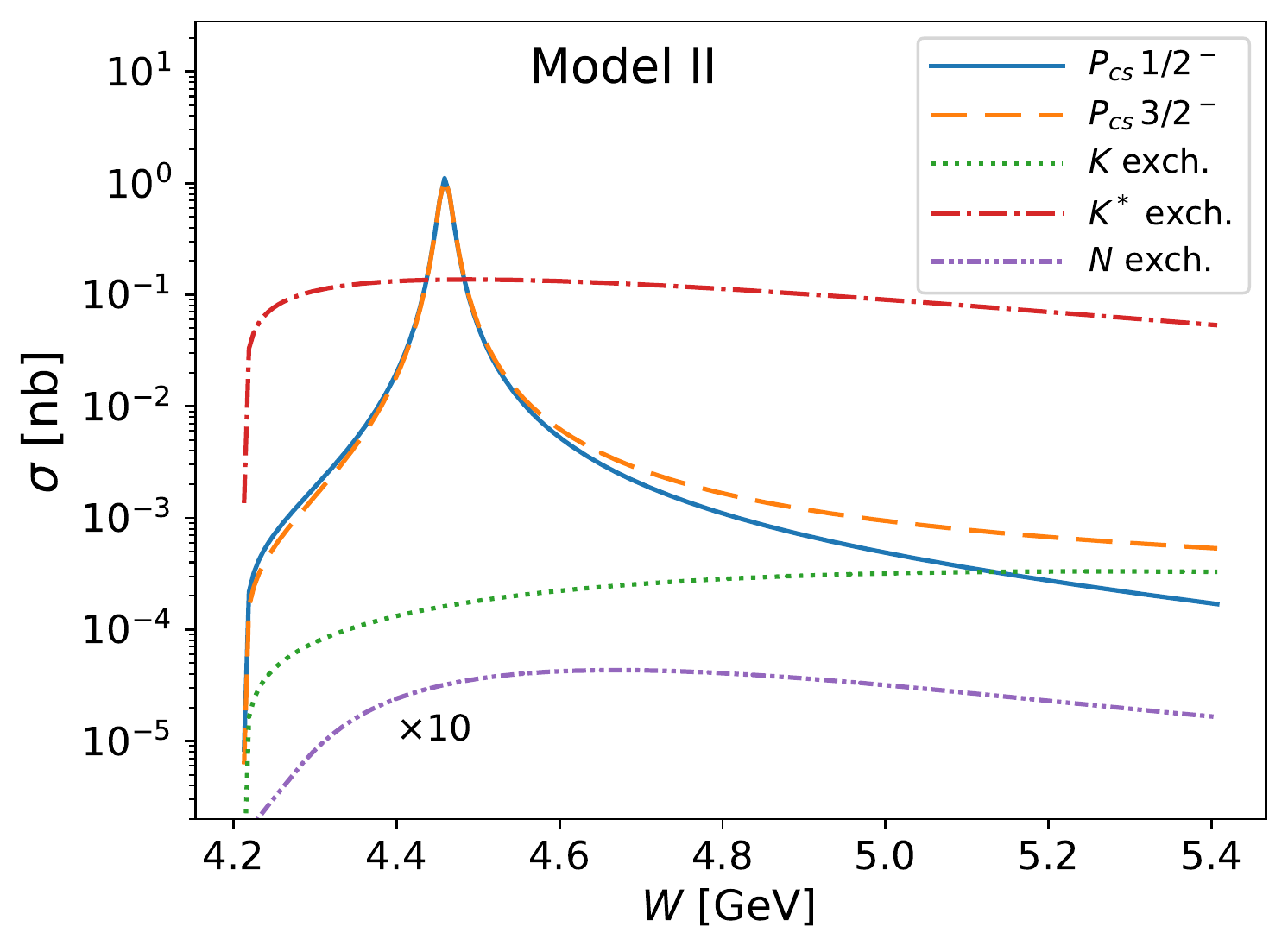}
\caption{Numerical results for the total cross section as a function
  of the total CM energy ($W$) from Model II. Notations are the same
  as in Fig.~\ref{fig:5}.
}
\label{fig:6}
\end{figure}
In Fig.~\ref{fig:6}, we draw the results for each contribution, which
are obtained from Model II, i.e., from the Regge approach. Since the
$s$-channel diagram is simply the same as that from Model I, we
discuss the contributions from $K^*$, $K$, and $N$ exchanges. As
mentioned previously, the value of the energy-scale parameter $s_0$ is
important for the size of the transition amplitudes. Since we use the
results from Model I as a guiding principle for determining $s_0$, we
expect that the magnitudes of the $K^*$- and $K$-Reggeon contributions
should be comparable to those from Model I. However, $s_0$ in the
Regge transition amplitude for $N$-Reggeon exchange in the $u$ channel
is constrained by the convergence condition. This means that the effect
of $N$ exchange is extremely small, so that we can even ignore it. 
Comparing the results from Model II, we find that the $K^*$
contributions from Model I exhibit different dependence on $W$. It is 
known from the asymptotic behavior of the differential cross sections
shown in Eq.~\eqref{eq:dcs} that the contributions of $K$- and
$K^*$-Reggeon exchanges should fall off slowly as $W$ increases. As
depicted in Fig.~\ref{fig:6}, $K^*$-Reggeon contribution indeed
decreases as $W$ increases. On the other hand, the results for $K^*$
exchange in Model I slowly increase as $W$ increases. This implies
that the effective Lagrangian method is limited in describing hadronic 
processes at higher energies, though it is a very effective
method in the vicinity of the threshold. 
The contribution of $K$-Reggeon exchange seems to arise as $W$
increases. However, if one further increases $W$, the contribution of
$K$-Reggeon exchange starts to fall off.

\begin{figure}[htp]
\includegraphics[scale=0.53]{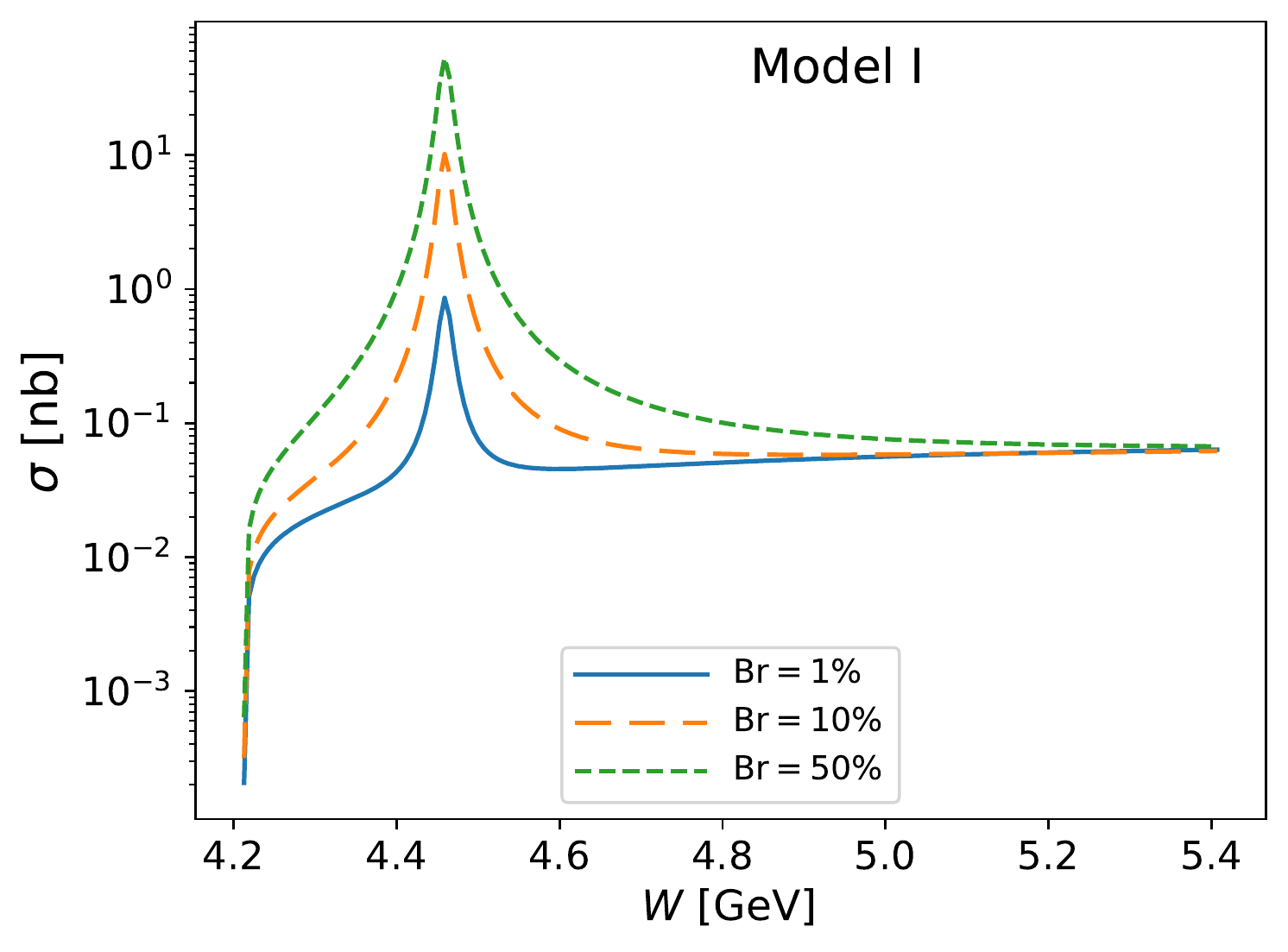}\hspace{0.7 cm}
\includegraphics[scale=0.53]{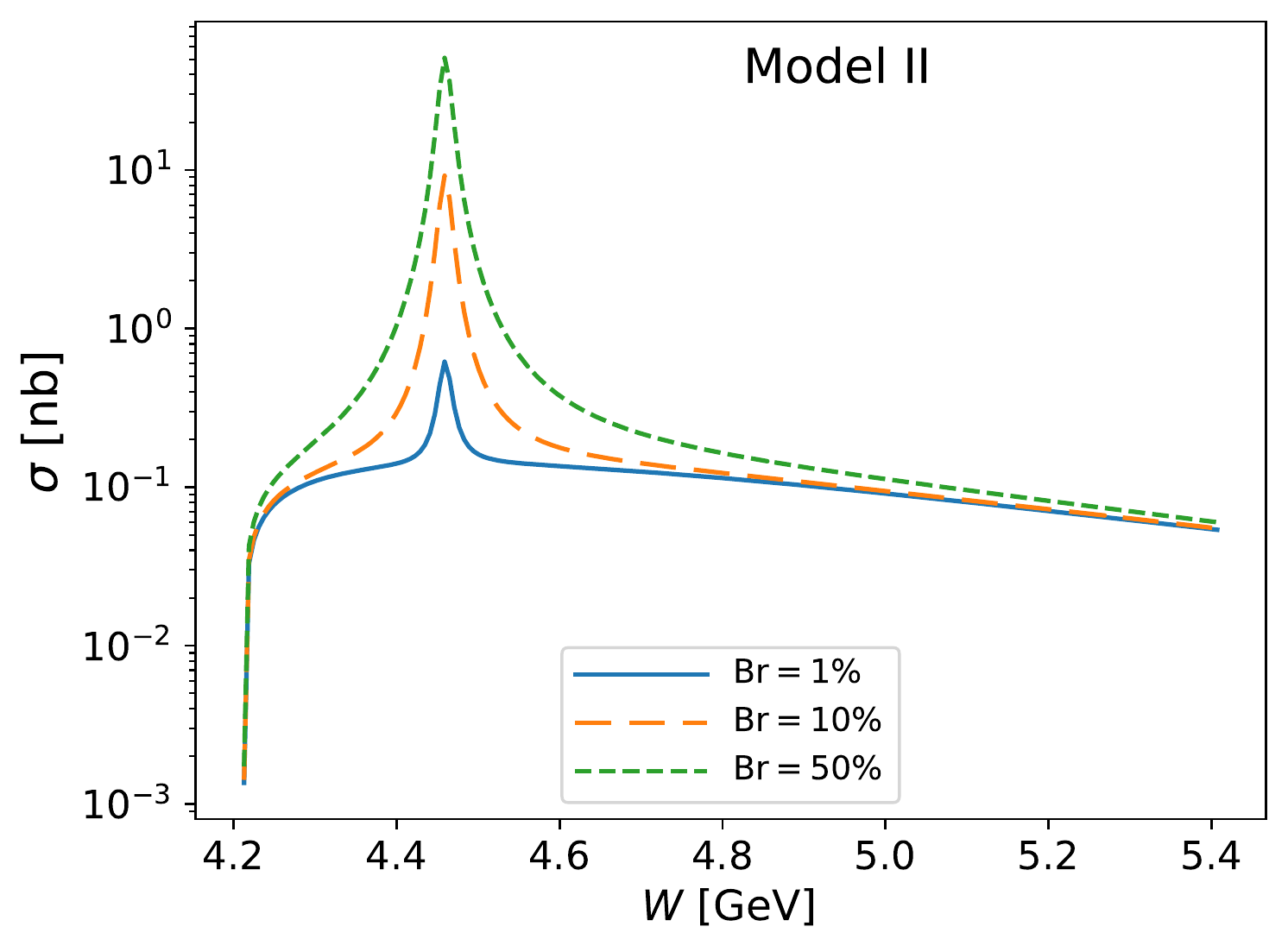}
\caption{Numerical results for the total cross section as a function 
of the total CM energy ($W$) from Model I (left panel) and Model II 
(right panel) with the branching 
ratio $B(P_{cs}\to J/\psi \Lambda)$ varied in the range of
$(1-50)\,\%$.} 
\label{fig:6s}
\end{figure}

In Fig.~\ref{fig:6s}, we will examine the dependence of the results
for the total cross section on the values of the branching ratio
of the $P_{cs}\to J/\psi \Lambda$ decay. As expected, if the value of
the branching ratio increases, the peak corresponding to $P_{cs}$
is enhanced clearly. Interestingly, the size of the peak reaches
approximately 10 nb when $\mathrm{Br}(P_{cs}\to J/\psi \Lambda)
=10~\%$ is used. When $\mathrm{Br}(P_{cs}\to J/\psi \Lambda)
=50~\%$ is employed, $\sigma$ is obtained to be almost 100 nb in the
vicinity of the resonance. This implies that if $\mathrm{Br}(P_{cs}\to
J/\psi \Lambda)$ is larger than $10~\%$, then $P_{cs}$ would have been
already found in the data for $K^-p$ scattering. Thus, the $1~\%$
branching ratio is a quite reasonable one, which is in agreement with
that from Refs.~\cite{Chen:2016qju, Xiao:2021rgp}.

\begin{figure}[htp]
\includegraphics[scale=0.7]{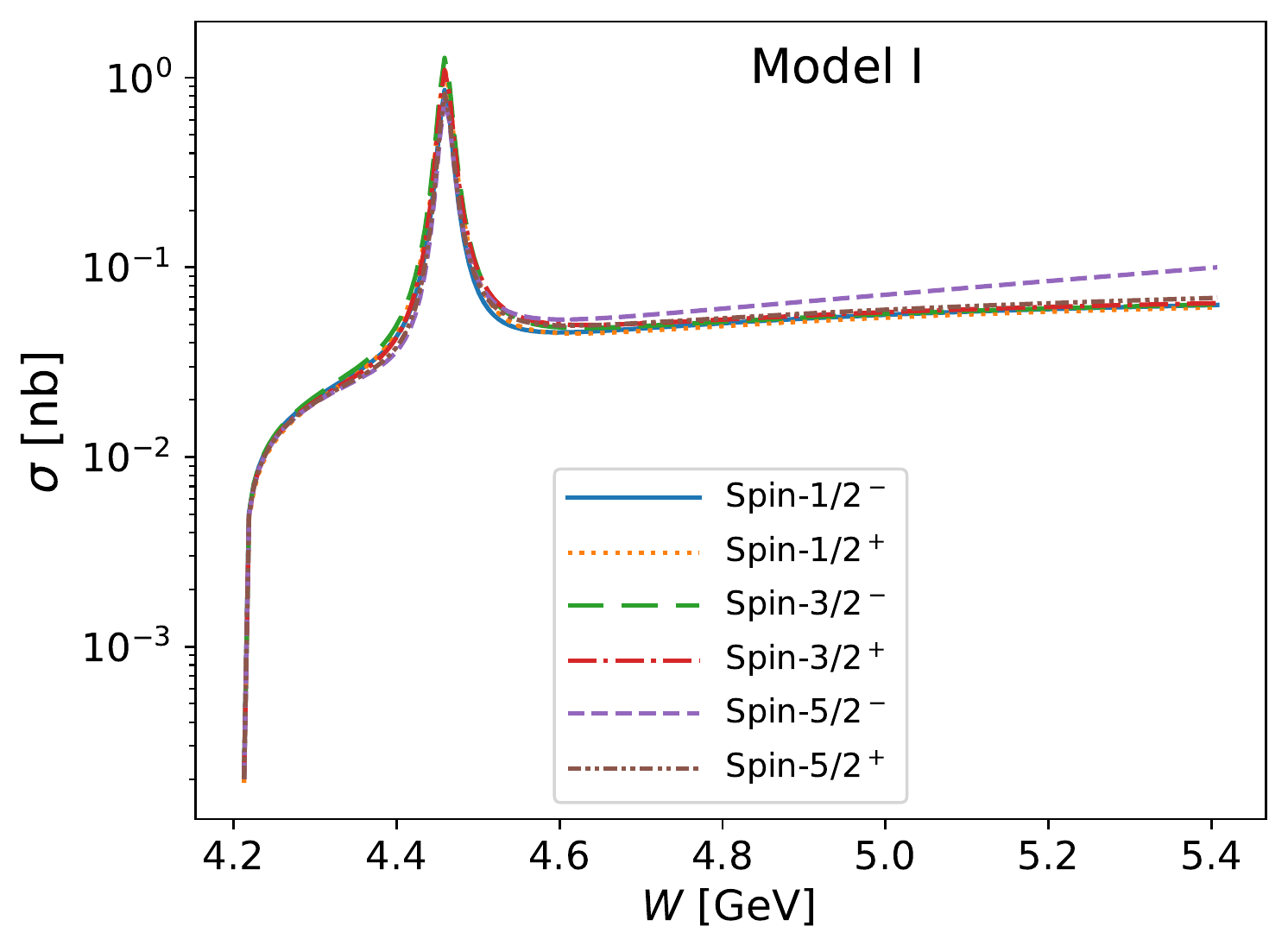}
\caption{Numerical results for the total cross section as a function
  of the total CM energy ($W$) from Model I with possible $J^P$ quantum
  numbers employed. Six different combinations of the spin and parity
  for the hidden-charm strange pentaquark $P_{cs}$ are taken into
  account. 
}
\label{fig:7}
\end{figure}
\begin{figure}[htp]
\includegraphics[scale=0.7]{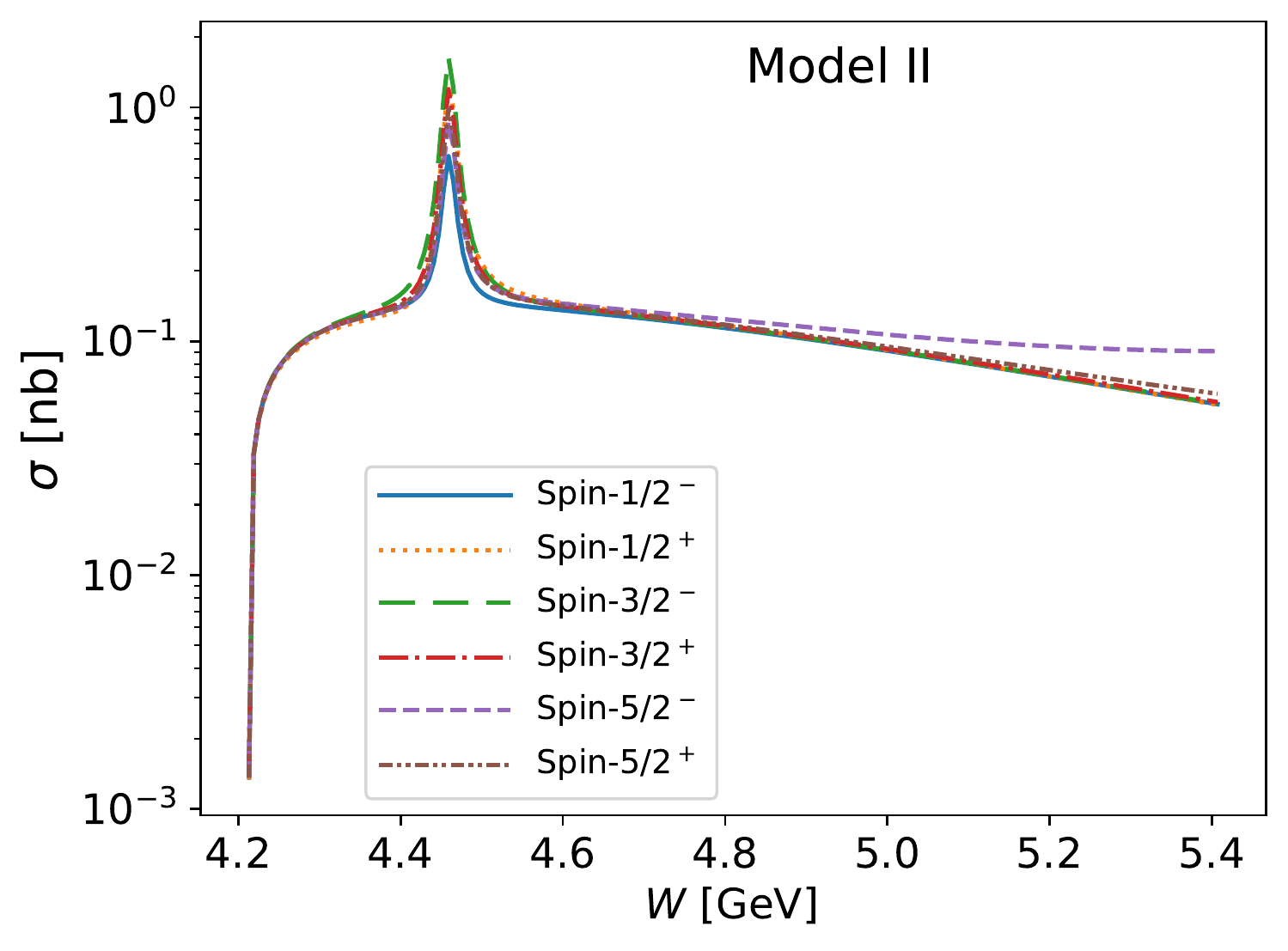}
\caption{Numerical results for the total cross section as a function
  of the total CM energy ($W$) from Model II with possible $J^P$ quantum
  numbers employed. Six different combinations of the spin and parity
  for the hidden-charm strange pentaquark $P_{cs}$ are taken into
  account. 
}
\label{fig:8}
\end{figure}

The spin-parity quantum number for $P_{cs}^0(4459)$ is experimentally
unknown yet. While it may have favorably either $J^P=1/2^-$ or
$J^P=3/2^-$, it is of great interest whether one can see how the
total cross sections and other observables for the $K^-p\to J/\psi
\Lambda$ reaction can provide a hint on the spin-parity quantum number
for $P_{cs}$. If the final states consisting of
$J/\psi$ and $\Lambda$ in the $S$ wave, the spin-parity quantum
numbers $J^P=1/2^-$ and $J^P=3/2^-$ of $P_{cs}$ will be
favored. However, there is no reason to reject other states with
higher values of the orbital angular momentum. Thus, we consider six
different combinations for the spin and parity for $P_{cs}$, i.e.,
$J^p=1/2^-$, $1/2^+$, $3/2^-$, $3/2^+$, $5/2^-$, and $5/2^+$. In
Figs.~\ref{fig:7} and~\ref{fig:8}, we draw the results for the total
cross sections by considering the six different combinations of the
spin-parity quantum numbers for $P_{cs}$, using Model I and Model II,
respectively. We find that except for the case of $J^P=5/2^-$, all the
results seem very similar each other. While the result for
$J^P=5/2^-$ from Model I shows a similar behavior in the resonance
region, it increases monotonically faster than all the other cases as
$W$ increases. On the other hand, the results from Model II decrease
monotonically as $W$ increases again except for the $J^P=5/2^-$
case. Even the total cross section for $P_{cs}(J^P=5/2^-)$  will
decrease, if $W$ further increases, though we did not show it in
Fig.~\ref{fig:8}. 

\begin{figure}[htp]
\includegraphics[scale=0.59]{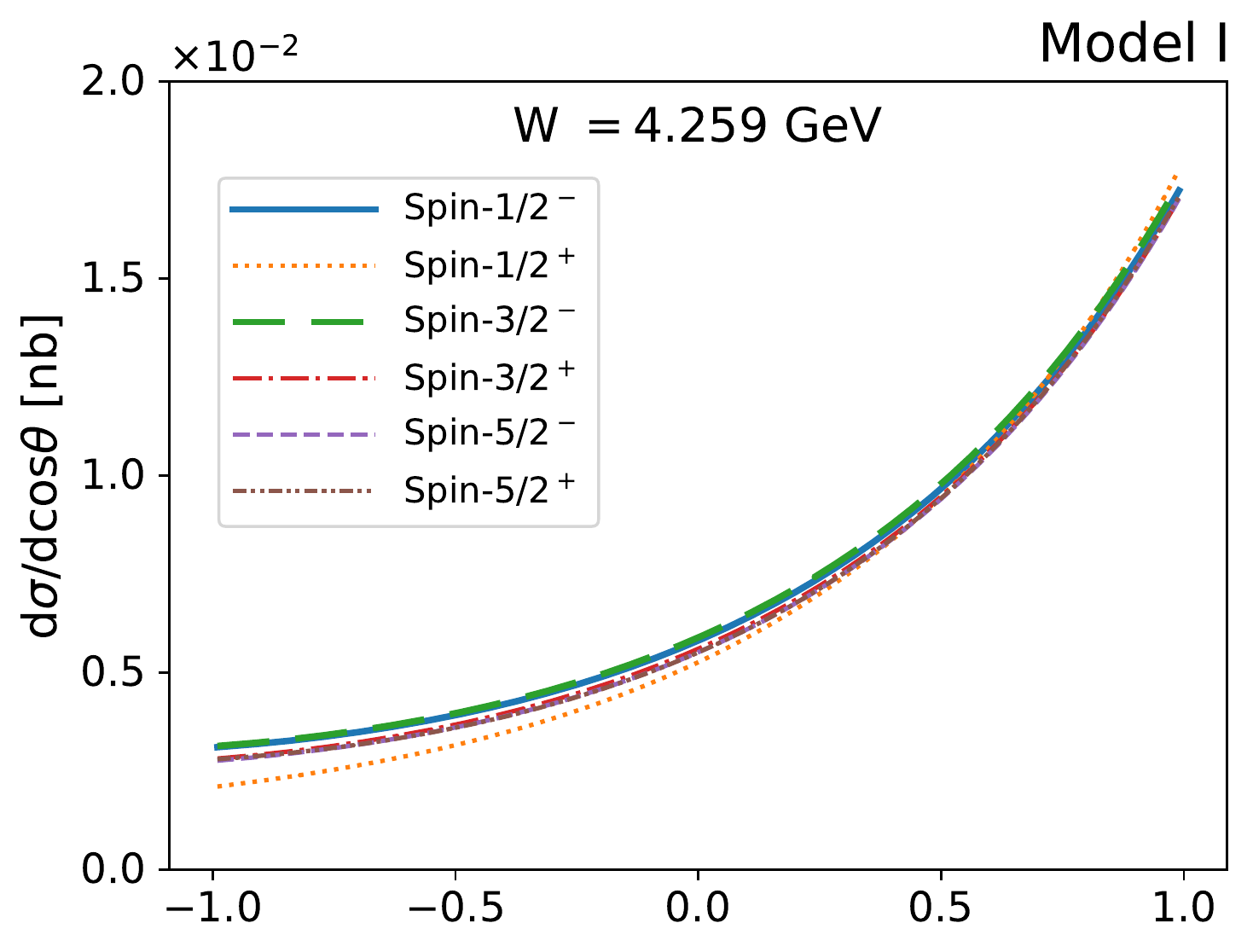}
\includegraphics[scale=0.59]{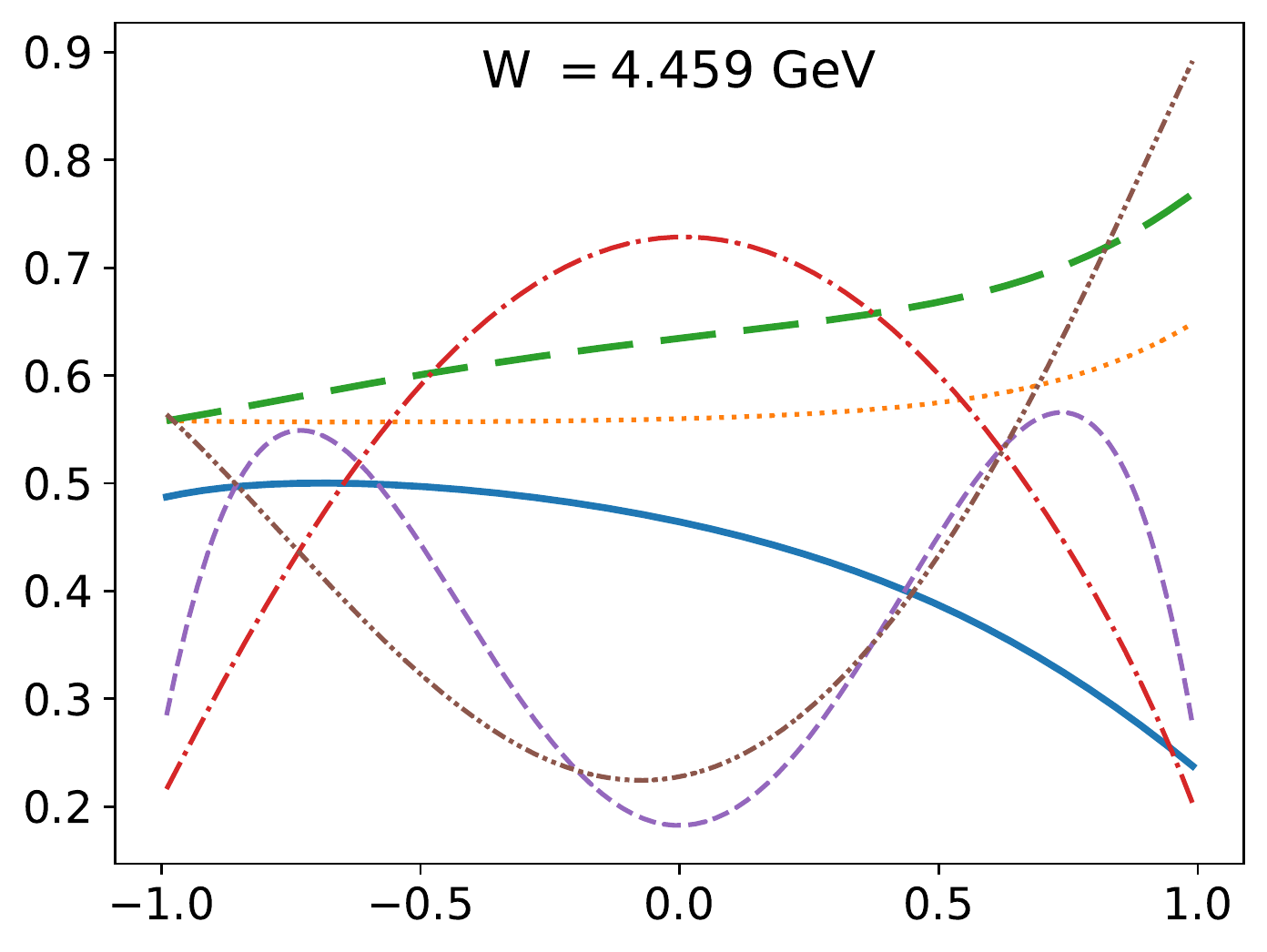}
\includegraphics[scale=0.59]{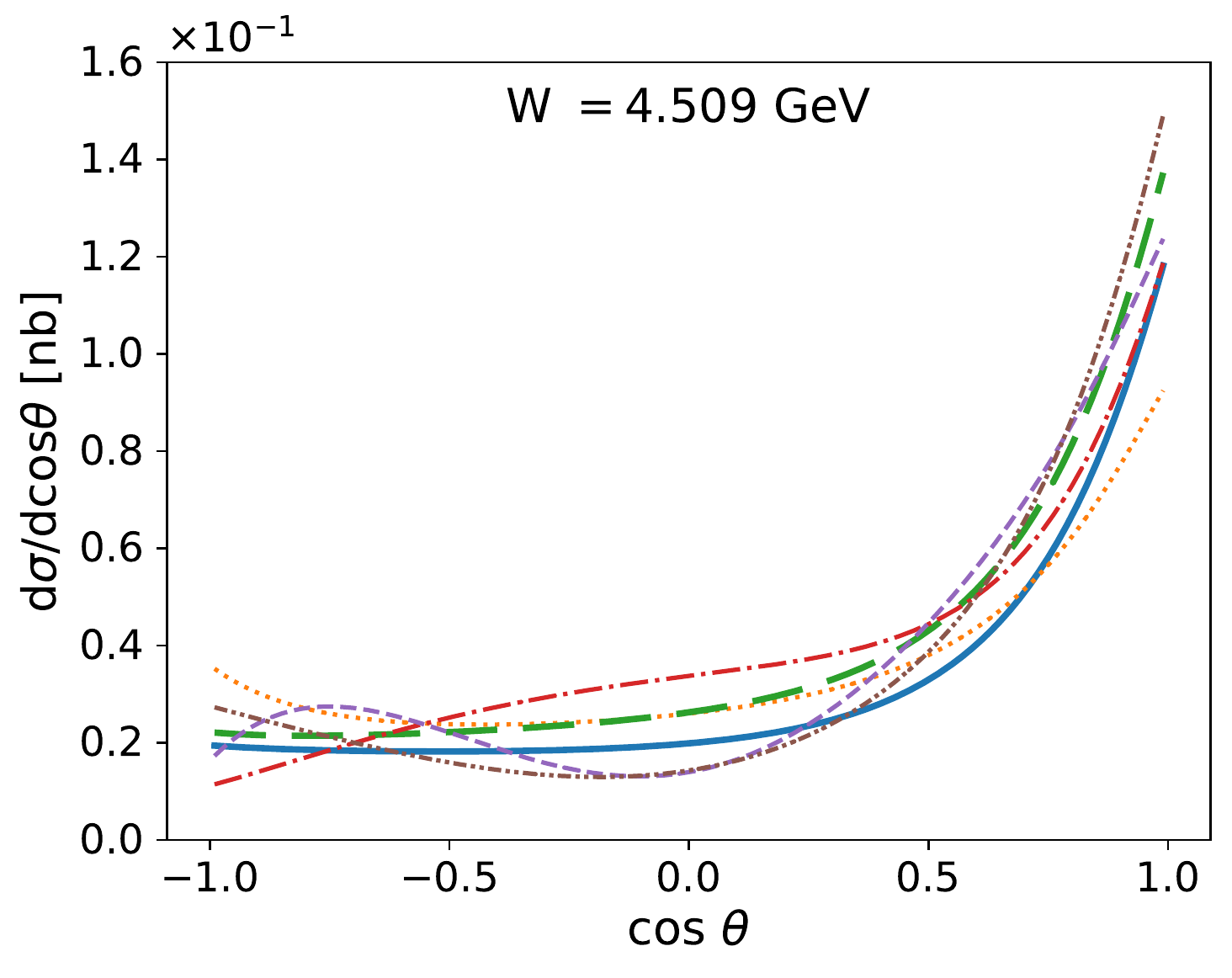}
\includegraphics[scale=0.59]{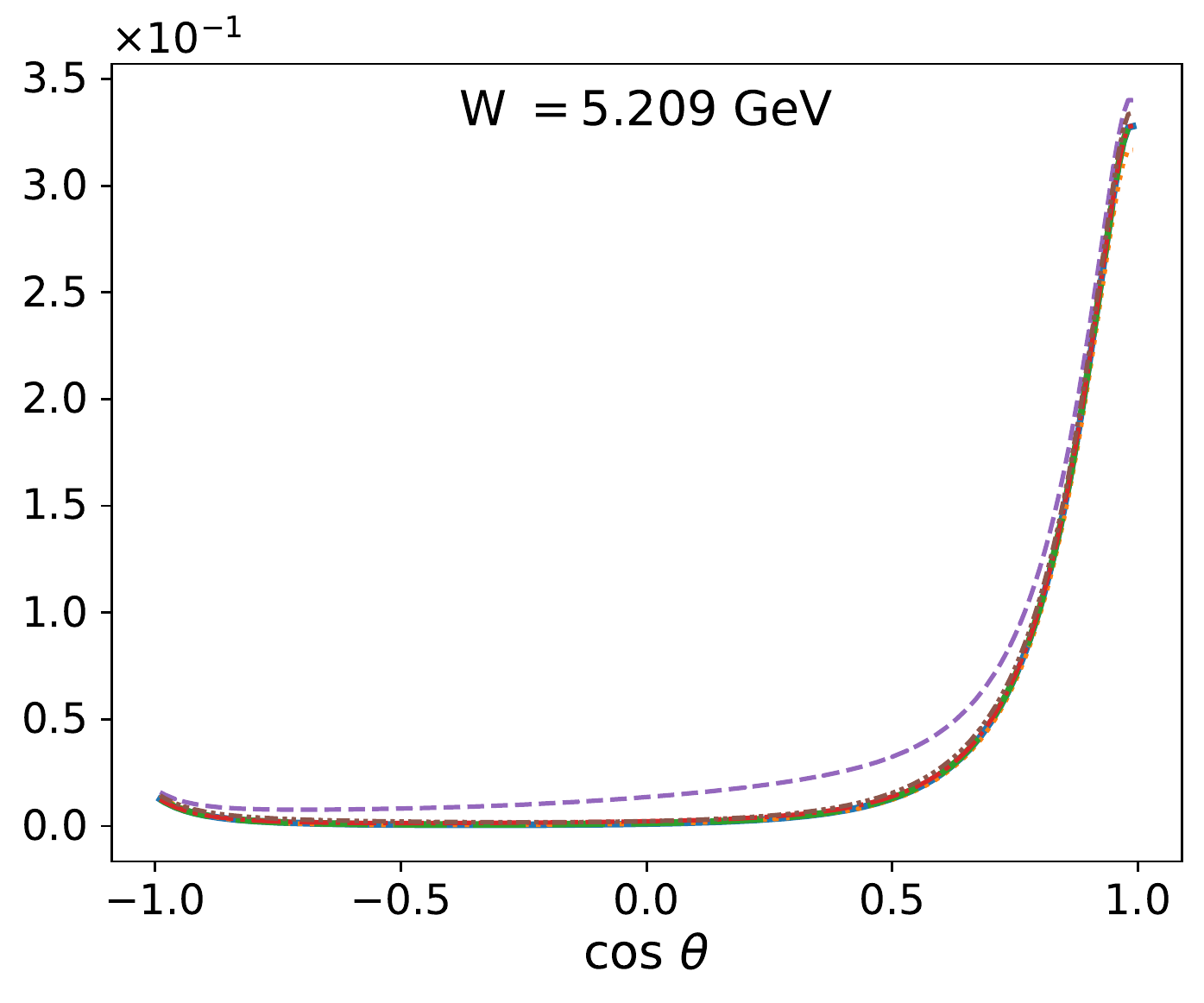}
\caption{Results for the differential cross sections
  ($d\sigma/d\cos\theta$) as functions of $\cos{\theta}$ for a given total
  energy ($W$) from Model I. The notation of the curves is the same
  as in Fig.~\ref{fig:7}.
}  
\label{fig:9}
\end{figure}
\begin{figure}[htp]
\includegraphics[scale=0.59]{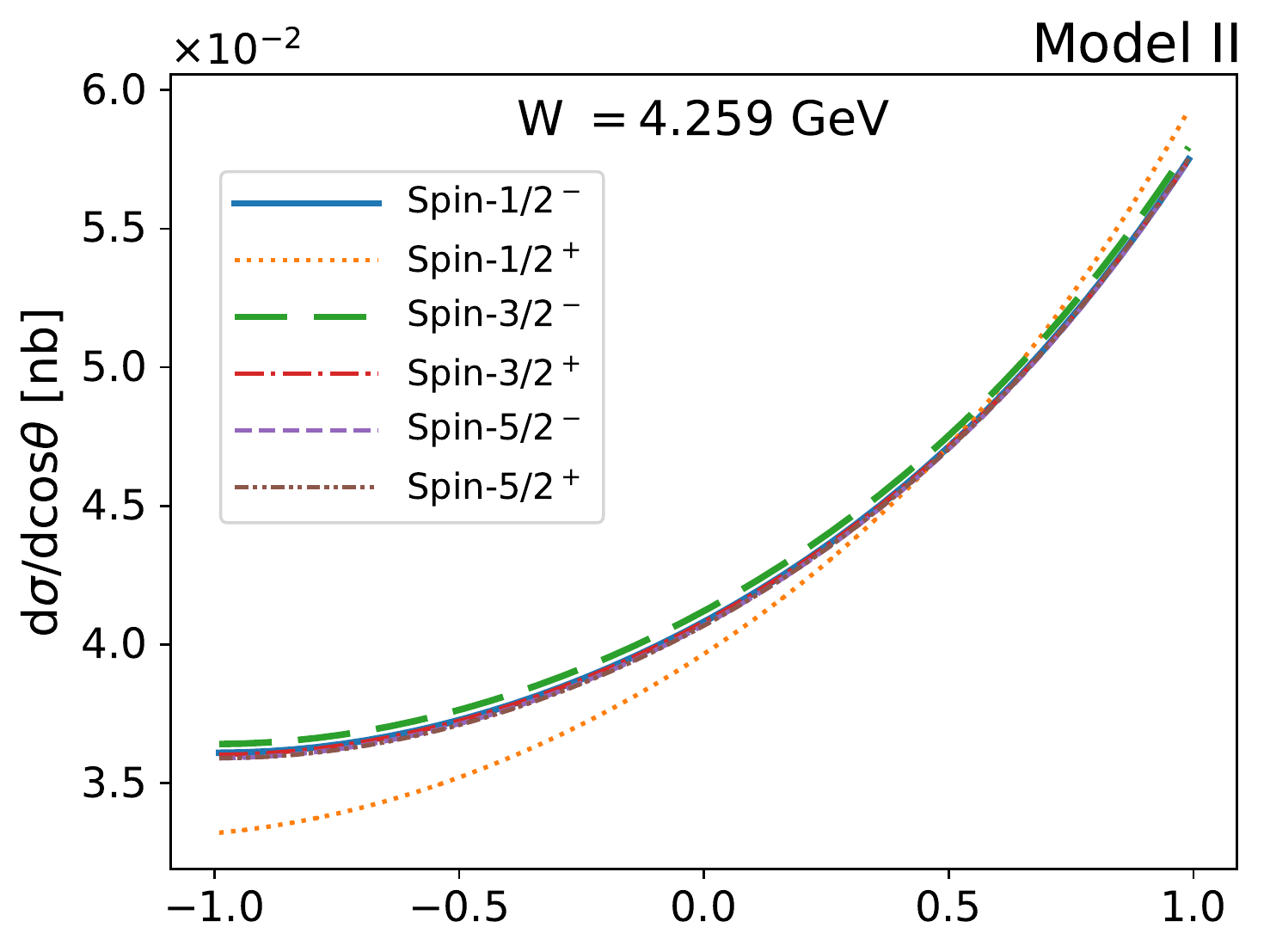}
\includegraphics[scale=0.59]{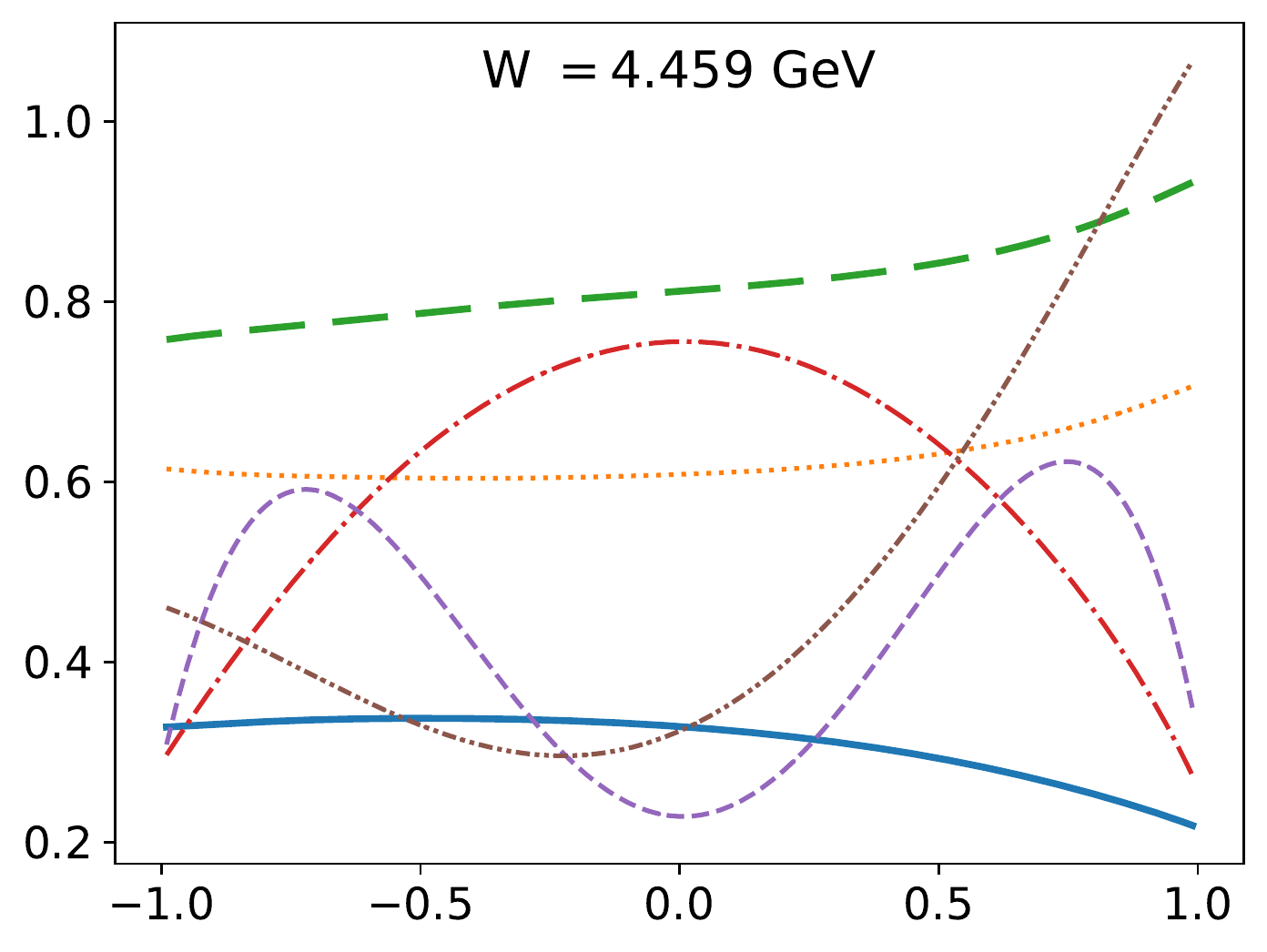}
\includegraphics[scale=0.59]{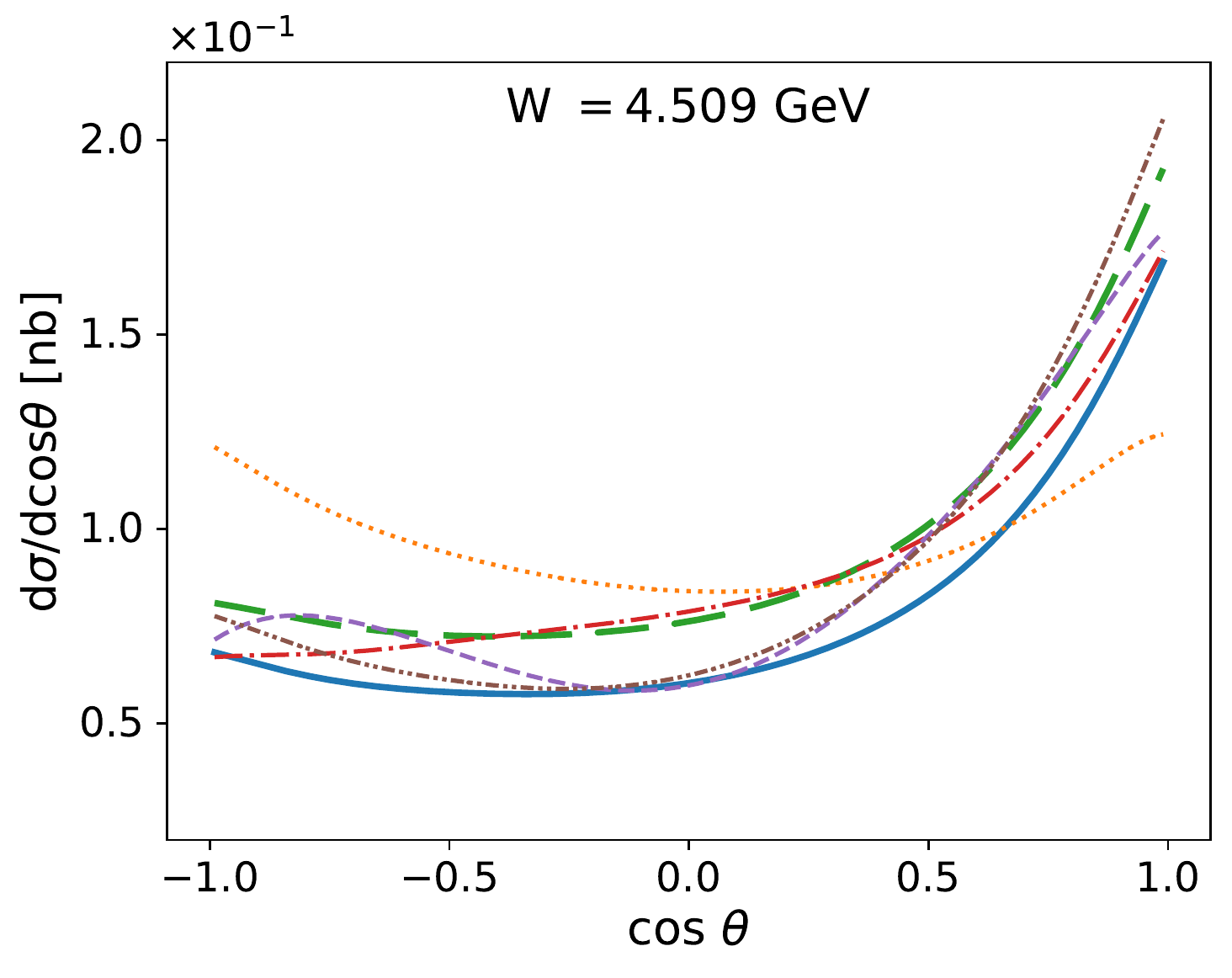}
\includegraphics[scale=0.59]{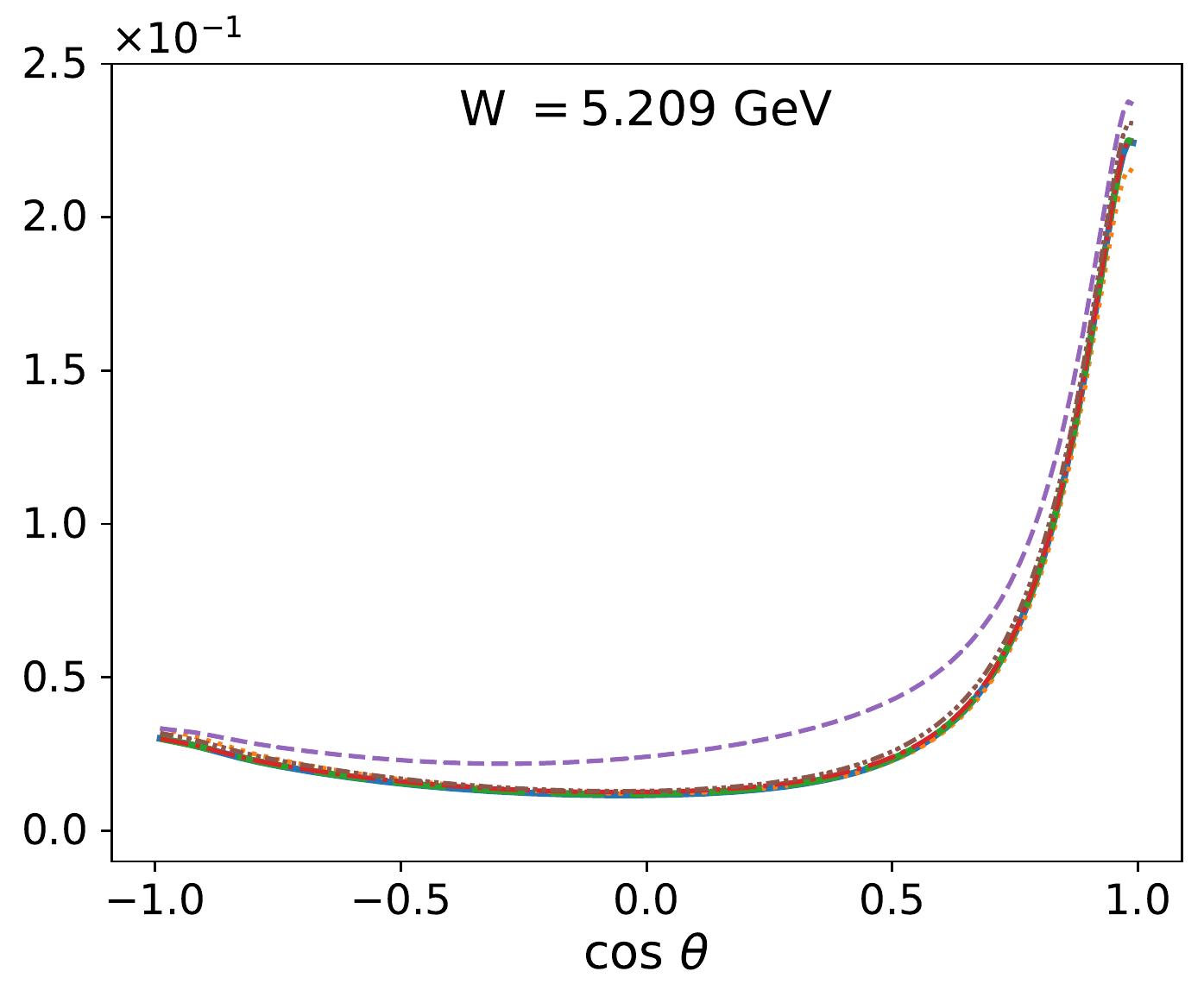}
\caption{Results for the differential cross sections
  ($d\sigma/d\cos\theta$) as functions of $\cos{\theta}$ for a given total
  energy ($W$) from Model II. The notation of the curves is the same
  as in Fig.~\ref{fig:7}.
}
\label{fig:10}
\end{figure}
Figures~\ref{fig:9} and~\ref{fig:10} depict the numerical results for
the differential cross sections $d\sigma/d\cos\theta$ as functions of
$\cos\theta$ with four different values of $W$ given. The results near
the threshold ($W=4.259$ GeV) clearly show that the magnitudes of the
differential cross sections in the forward direction are the largest
ones and then decrease monotonically as $\cos\theta$ goes from $+1$ to
$-1$. So, the results for the differential cross sections are mostly 
diminished in the backward direction. While the results from Model II
at $W=4.259$ GeV exhibit similar behaviors to those from Model I,
detailed dependences on $\cos\theta$ look different. 

When it comes to the resonance region at $W=4.459$ GeV, the
results are noticeably distinguished for different assignments of
$J^P$ to $P_{cs}$. Scrutinizing first the results for the cases of
$J^P=1/2^-$ and $J^P=3/2^-$, we find that the $\cos\theta$ dependence
of them is rather different. The result for $P_{cs}^0(J^P=1/2^-)$
is suppressed in the forward direction, whereas that for
$P_{cs}^0(J^P=3/2^-)$ gets enhanced as $\cos\theta$ increases. This
implies that the resonance and the $K^*$ exchange contributions
interfere differently each other. When $J^P=1/2^-$ is assumed, the two
terms interfere destructively, while they do constructively with
$J^P=3/2^-$ assumed. When one takes $J^P=3/2^+$ for $P_{cs}$, the
corresponding differential cross section becomes maximum at
$\cos\theta=0$, i.e., $\theta=90^\circ$. On the other hand, if
$J^P=5/2^+$ is assumed, the values of the differential cross section
will be the minimum at $\theta=90^\circ$. When $J^P=5/2^-$ is imposed,
the result for $d\sigma/d\cos\theta$ becomes more complicated. 
Thus, the measurement of the differential cross
sections near the resonance region may provide one way of determining
the spin-parity quantum numbers for $P_{cs}$ in the $K^-p\to J/\psi
\Lambda^0$ reaction. 

\begin{figure}[htp]
\includegraphics[scale=0.59]{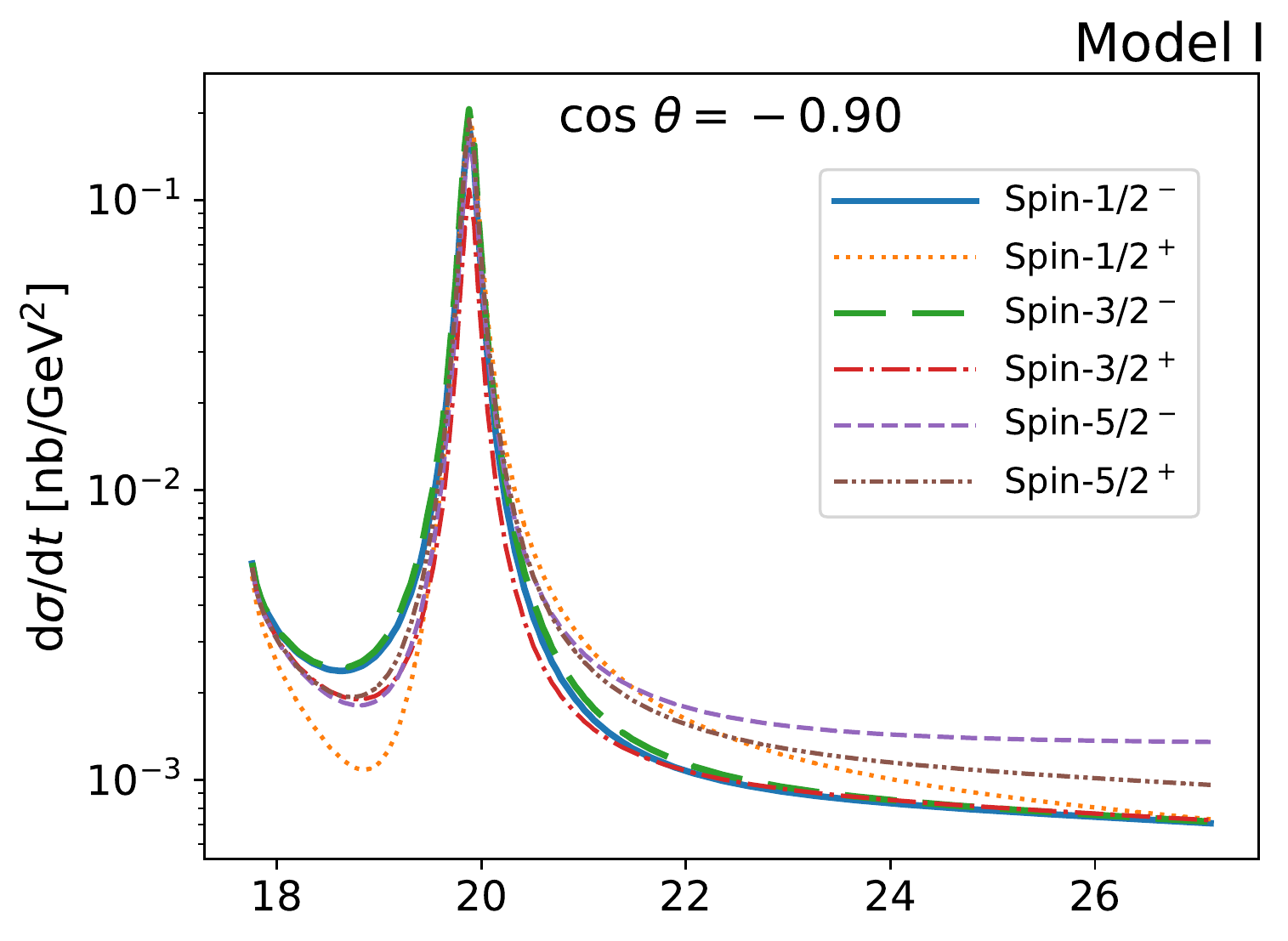}
\includegraphics[scale=0.59]{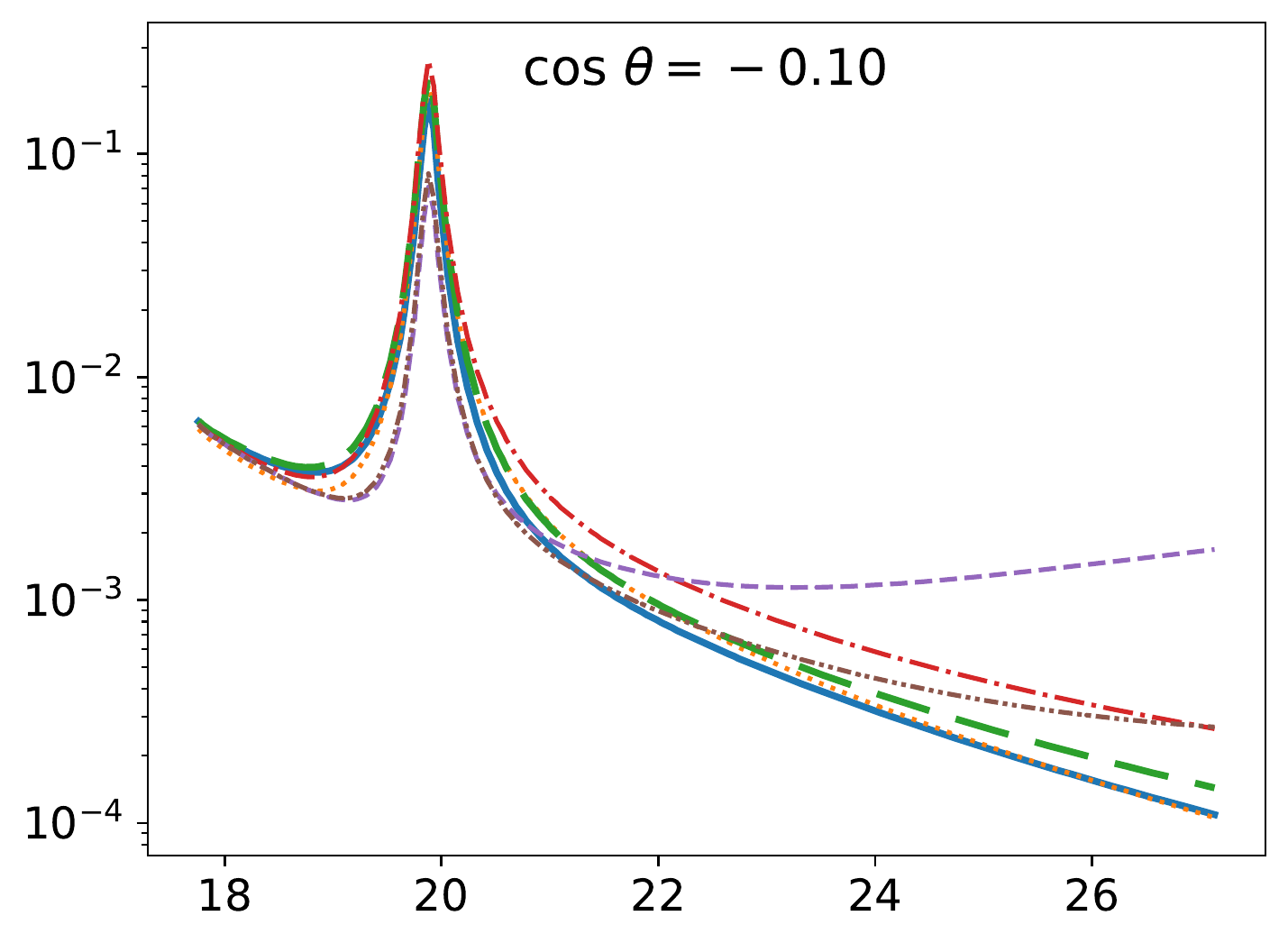}
\includegraphics[scale=0.59]{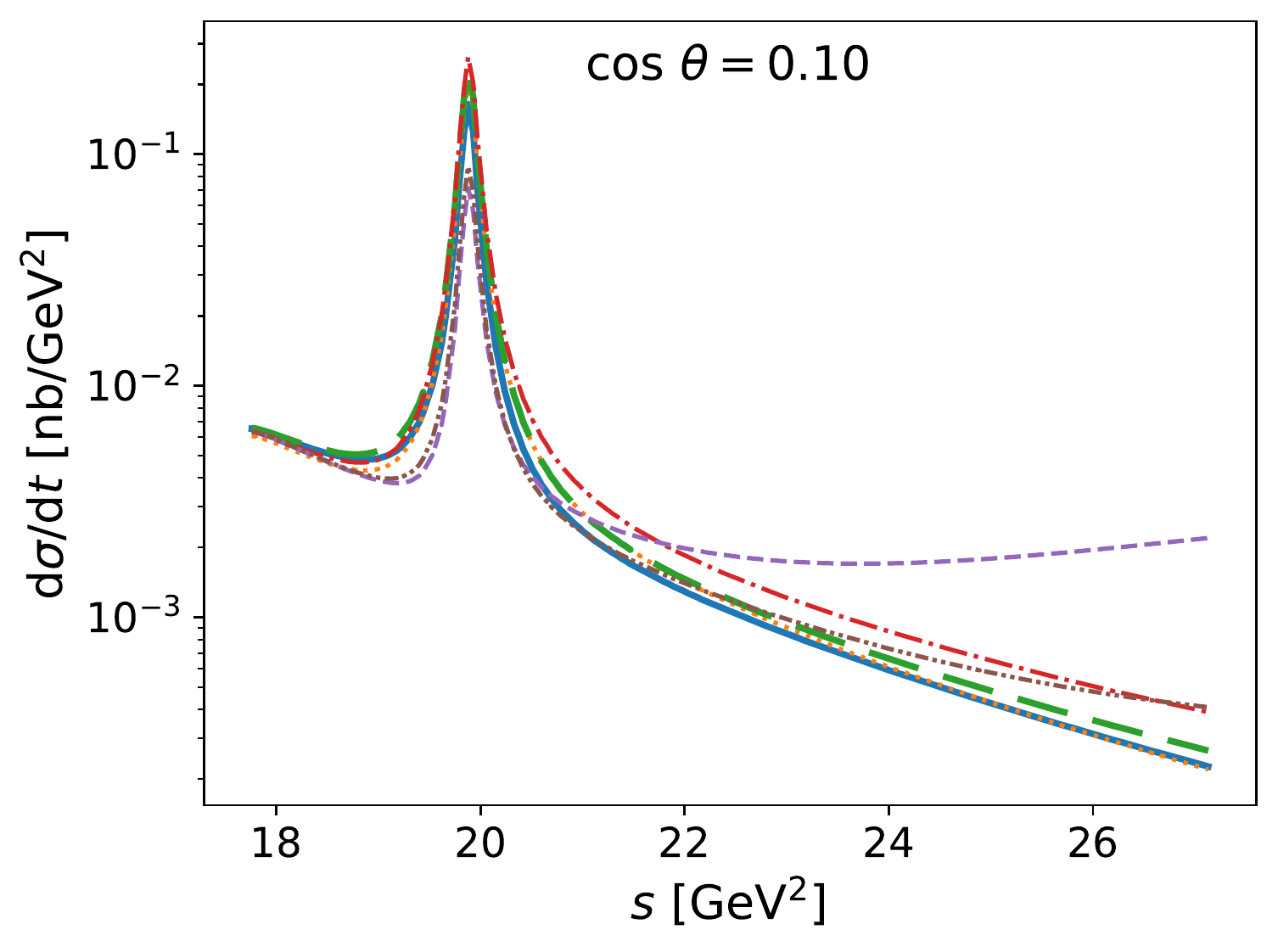}
\includegraphics[scale=0.59]{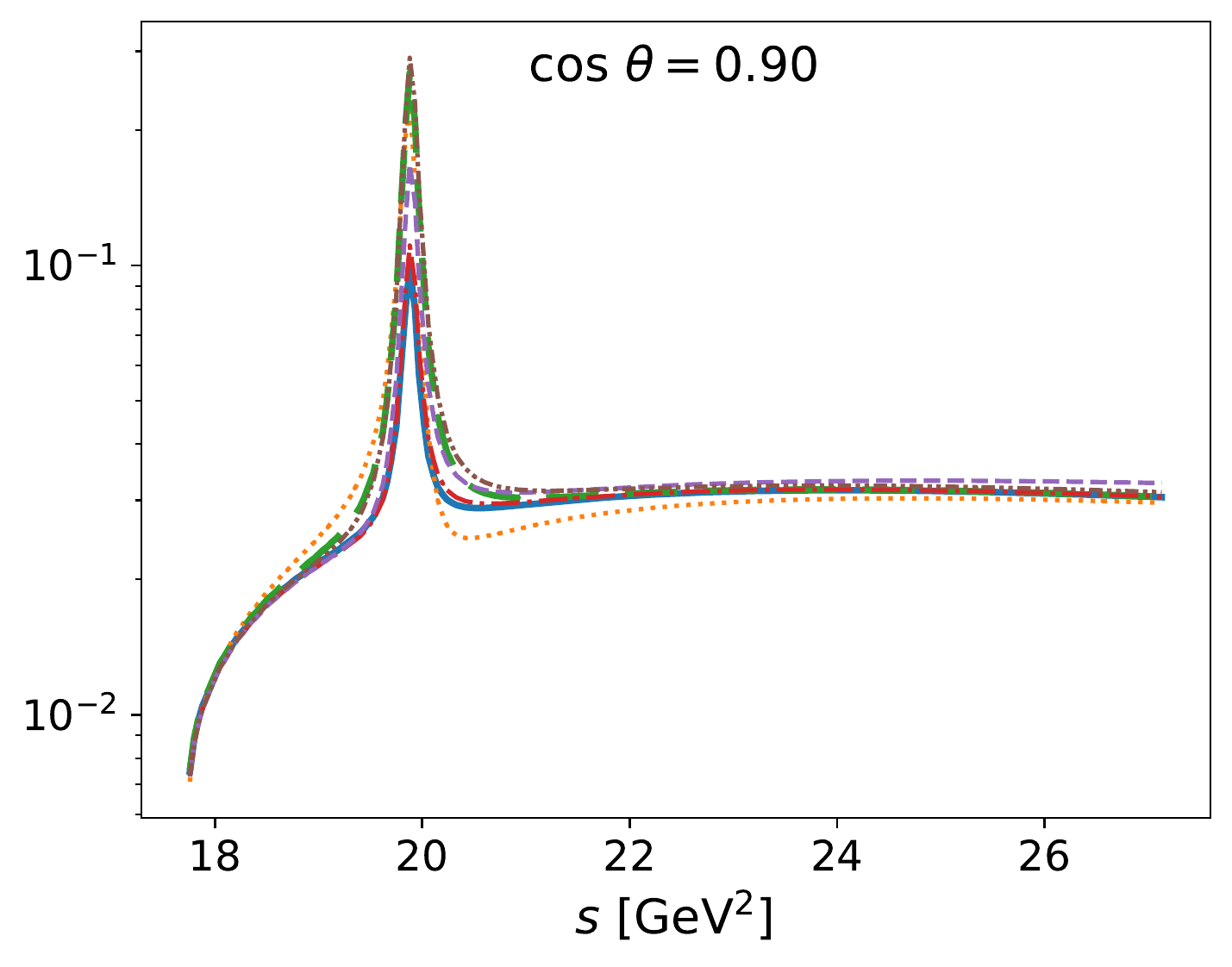}
\caption{Results for the differential cross sections
  ($d\sigma/d\cos\theta$) as functions of $s$ for a given angle
  ($\cos\theta$) from Model I. The notation of the curves is the same
  as in Fig.~\ref{fig:7}.
}
\label{fig:11}
\end{figure}
\begin{figure}[htp]
\includegraphics[scale=0.59]{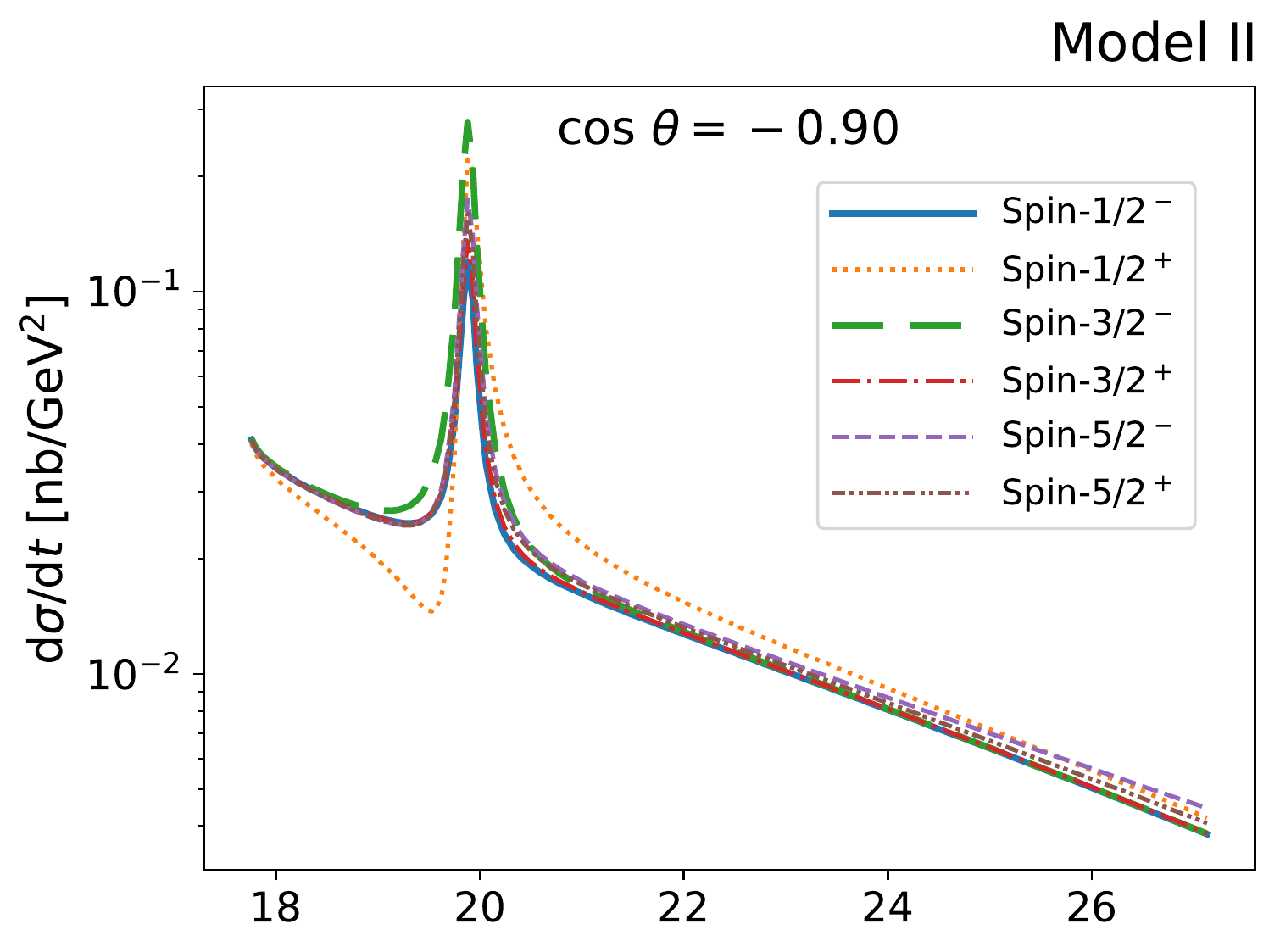}
\includegraphics[scale=0.59]{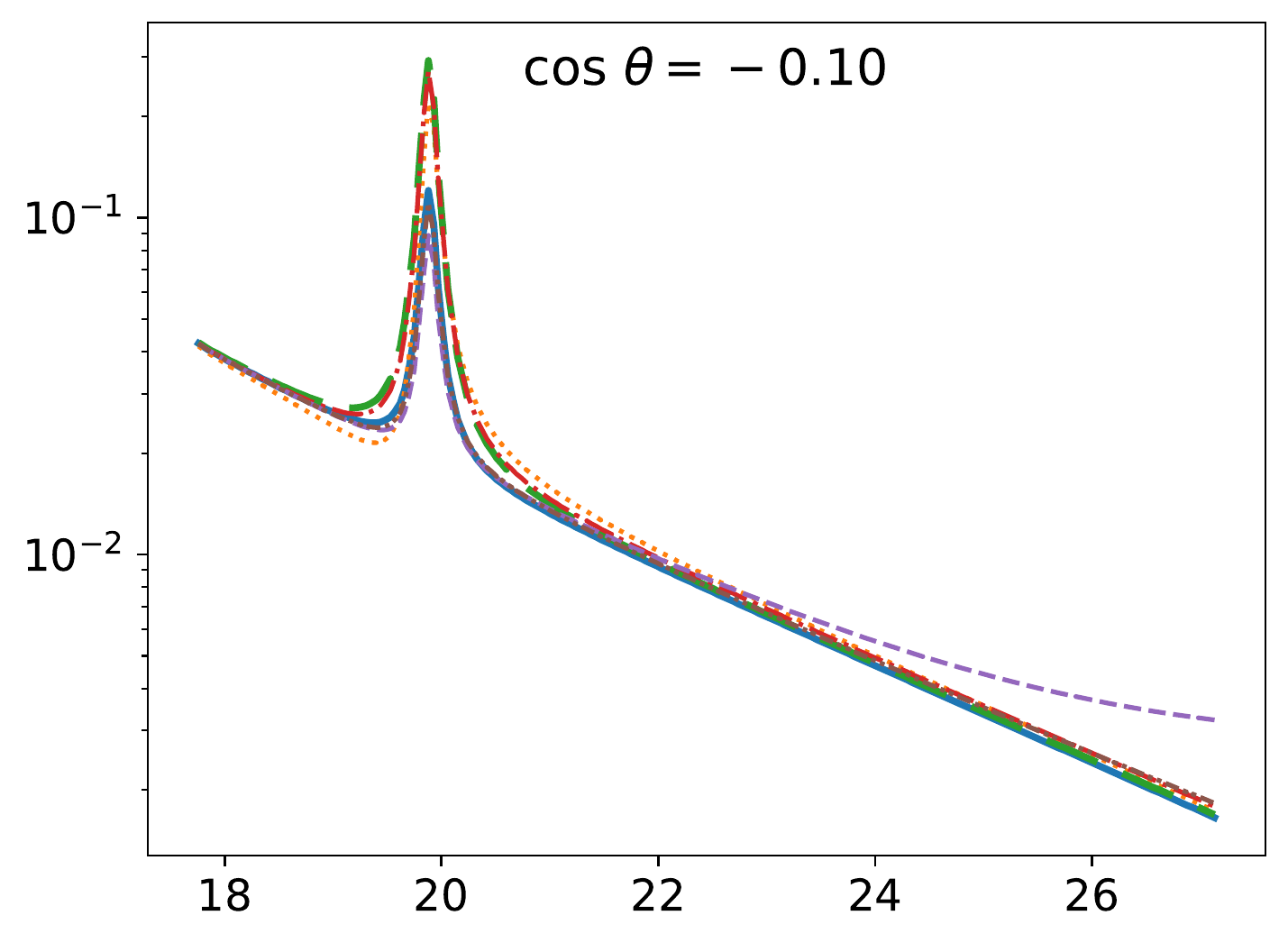}
\includegraphics[scale=0.59]{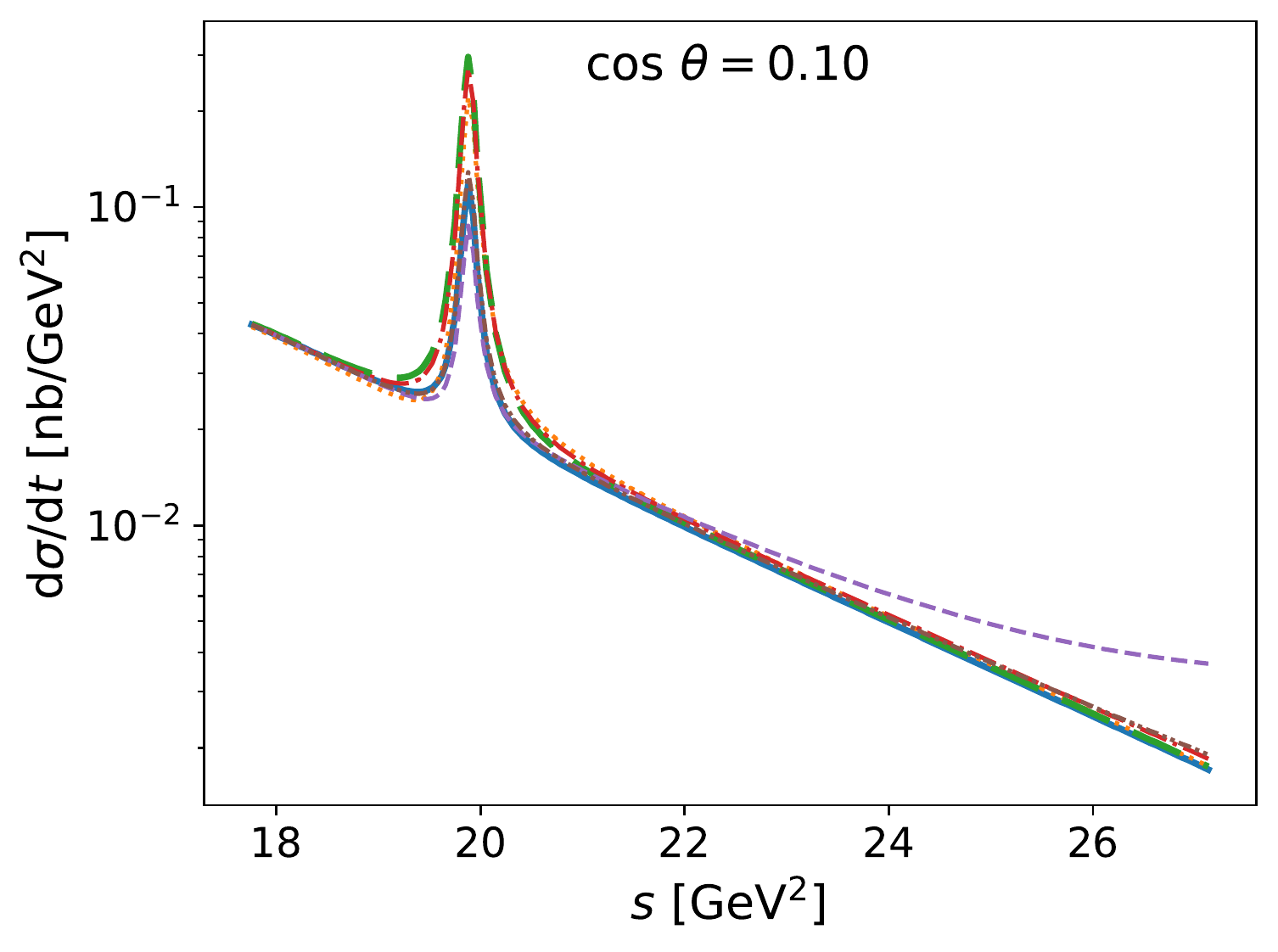}
\includegraphics[scale=0.59]{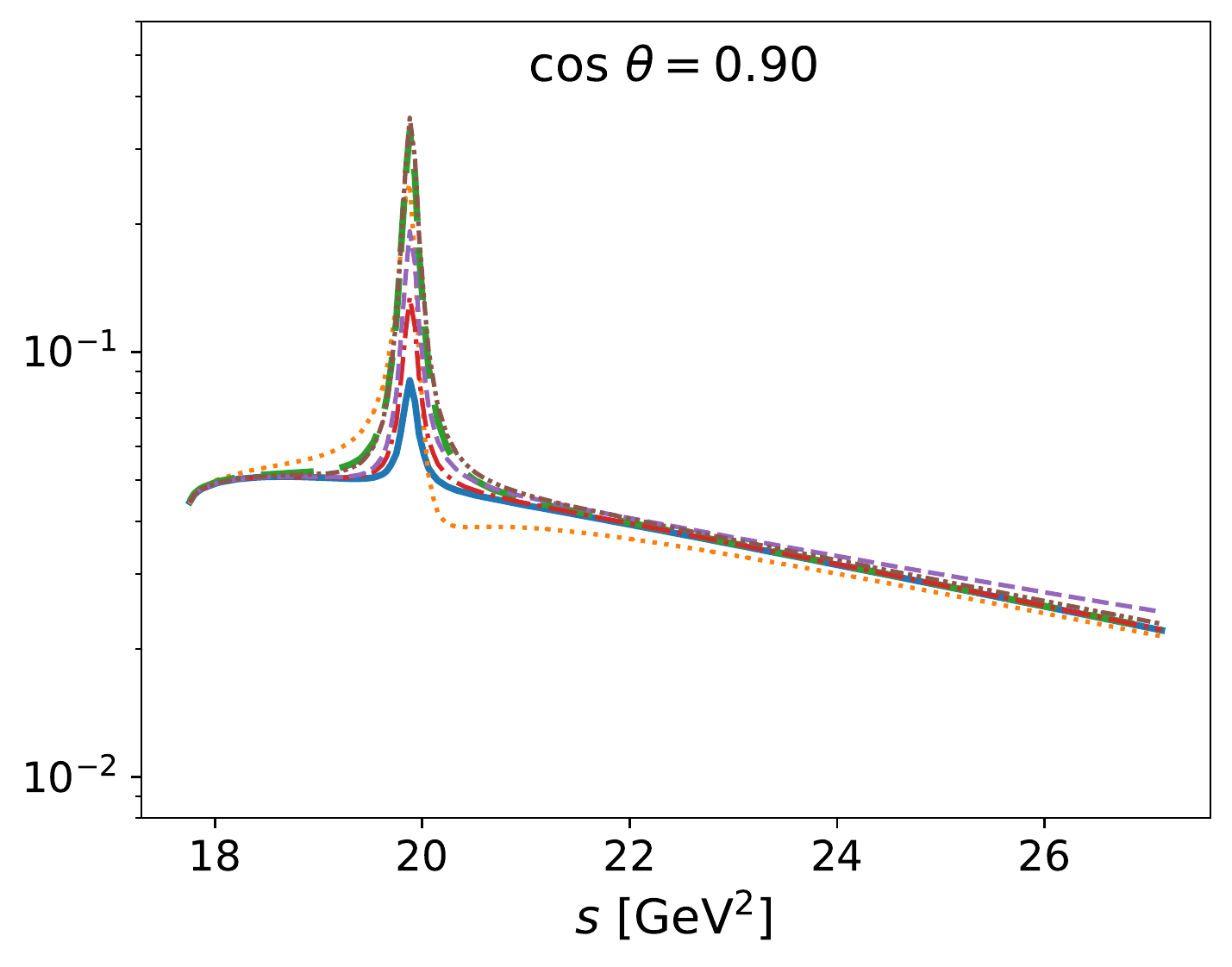}
\caption{Results for the differential cross sections
  ($d\sigma/d\cos\theta$) as functions of $s$ for a given angle
  ($\cos\theta$) from Model II. The notation of the curves is the same
  as in Fig.~\ref{fig:7}.
}
\label{fig:12}
\end{figure}
In Figs.~\ref{fig:11} and~\ref{fig:12}, we depict the results for the
differential cross sections $d\sigma/dt$ as functions of $W$, which
are obtained from Model I and II, respectively, varying the scattering
angle $\cos\theta$ from $\cos\theta =0.9$ to $\cos\theta=-0.9$. In
the forward direction, the results for $d\sigma/dt$ look similar to
those for the total cross sections. However, the results at
$\cos\theta =0.1$ and $\cos\theta=-0.1$ enable us to distinguish among
those with different $J^P$. While the shapes of the resonances
corresponding to $P_{cs}$ look all similar, one can distinguish them
each other as $s$ increases. Getting out of the resonance regions, the
results decrease as $s$ increases except for the case of $J^P=5/2^-$,
in particular, when one uses Model I. In fact, 
we already found this behavior in the results for the total
differential cross sections. However, we can see this particular
behavior more prominently in those for $d\sigma/dt$ as $s$
increases. The reason is clear. As shown in
Eqs.~\eqref{eq:8},~\eqref{eq:9}, and~ \eqref{eq:10}, the 
transition amplitudes contain strong momentum dependence with higher
spin of $P_{cs}$ assumed. As $s$ further increases, the results for
$J^P=5/2^+$ also start to increase slowly. This comes from the fact
that the difference in the parity also affects the interference
effects. Moreover, this peculiar dependence of $d\sigma/dt$ for 
$J^P=5/2^-$ on $s$ implies that the effective Lagrangian method may
not be valid anymore at very high energies. On the other hand, the
Regge approach nicely produces the asymptotic behavior of $d\sigma/dt$
as $s$ increases. Even the result for $d\sigma/dt$ with $J^P=5/2^-$
assigned starts to fall off when $s$ further increases, though we do
not show in Fig.~\ref{fig:12} explicitly.

\section{Summary and conclusion}
In the present work, we aimed at investigating the production of
$P_{cs}^0(4459)$ in the $K^- p\to J/\psi \Lambda^0$ reaction,
employing two different theoretical frameworks, i.e. the effective
Lagrangian method and the Regge approach. We call these two different
approaches as Model I and Model II, respectively. We first determined
the coupling constants for all the relevant hadronic vertices. Since
there is lack of experimental data on them, we made various reasonable
assumptions. To determine the coupling constant for the $P_{cs} J/\psi
\Lambda$ vertex, we assumed that the branching ratio of $P_{cs}\to
J/\psi \Lambda$ decay is about $1~\%$. That of 
$P_c\to J/\psi N$ was also proposed to be about $0.01~\%$. Thus, the
coupling constant $g_{P_{cs} J/\psi \Lambda}$ is of order $0.1$. When 
one considers the hidden-charm pentaquark with higher spins ($J^P\ge
3/2^{\pm}$), the tensor couplings are naturally introduced. However,
since $J/\psi$ is an isosinglet, the tensor coupling constants can be
neglected as in the case of the $\omega$ meson. Moreover, since we are
mainly interested in the resonance $P_{cs}$ region, which is not far
from the threshold of $J/\psi$ and $\Lambda$, the contributions from
the tensor couplings can be taken to be very small. Since the 
Okubo-Zweig-Iizuka rule indicates that the coupling between a
nucleon and a $\phi$ meson ($s\bar{s}$) should be very small, the
same is applied to the coupling between a hyperon and a charmonium
($c\bar{c}$). Thus, we also took the value of the coupling
constant for the $KP_{cs}N$ vertex to be very small. By
estimating the branching ratio of $P_{cs}\to K^- p$, we found that the
value of the $KP_{cs}N$ coupling constant is of order
$10^{-3}$. 

Since the decay widths of $J/\psi$ to the $K$ and $K^*$mesons
are known, we were able to determine directly the corresponding
coupling constants from experimental data. 
Our results are obtained by setting the cut-off mass for the
off-shell $P_{cs}$ to $5$ GeV, which is a rather plausible choice, 
even if we observe that our predictions are extremely sensitive to 
the value of the cut-off, which means that if one can change the 
value a little bit, then the results would be very much changed. 
On the other hand, since there are no experimental data to determine 
the values of the cut-off masses certain uncertainties caused by them 
are inevitable. As for the form factors for $K$ and $K^*$, we fixed
the values of the cutoff masses to be $\Lambda_K=1$ GeV and
$\Lambda_{K^*}=1.4$ GeV. In the case of the Regge approach, we
considered a nonlinear form of the $K$ and $K^*$ Regge trajectories,
which fit the experimental data much better than the linear ones.  

We first scrutinized the results for the total cross sections as
functions of the CM total energy $W$, with different spin-parity
quantum numbers $J^P$ taken into account. While the shape of the
resonance does not much depend on the given value of $J^P$, the
dependence on $W$ is different. In particular, the result with
$J^P=5/2^-$ increases faster than the other ones as $W$ increases. We
found a similar feature in the case of the Regge approach. However,
$W$ increases further, all the results for the total cross sections
are lessened as $W$ increases. Thus, the Regge approach produces the
results more consistently than those from the effective Lagrangian
method. Secondly, we examined the results for the differential cross
sections as functions of the scattering angle with several different
values of the CM total energy. The results in the resonance region
clearly are distinguished, as different sets of the spin-parity
quantum numbers are used. This implies that the measurement of
differential cross sections for the $K^-p\to J/\psi \Lambda$ reaction
may give a clue on the spin-parity quantum number of $P_{cs}$. We also
studied the differential cross sections $d\sigma/dt$ as functions of
the CM total energy squared, i.e., $s$. When the scattering angle near
$\theta=90^\circ$, $s$ dependences of the differential cross sections
prominently reveal the differences among the results with different
sets of $J^P$. 

The present results may be used as a theoretical guide for possible
future experiments for findings of the hidden-charm pentaquarks with
strangeness. Similar studies for other $P_{cs}$ are also under way. 
\begin{acknowledgments}
 The present work was supported by Basic Science Research Program
 through the National Research Foundation of Korea funded by the
 Ministry of Education, Science and Technology
 (Grant-No. 2018R1A5A1025563).  
\end{acknowledgments}


\end{document}